\documentclass{cernyrep}
\usepackage{texnames}
\usepackage[T1]{fontenc}
\usepackage{varwidth}
\usepackage{xcolor}

\usepackage[bookmarks, colorlinks=true, linktoc=page, pdftex, linkcolor=black, citecolor=black, urlcolor=blue]{hyperref}
\usepackage{fancyhdr}
\fancyhfoffset{4 mm}
\fancypagestyle{ARTTITLE}{%
\fancyhf{} 
\lhead{\small{CERN Accelerator School Proceedings ---
{\it Advanced Accelerator Physics} ---  Spa,  Belgium, 2024}}
\lfoot{\hspace{3mm} Available online at \url{https://cas.web.cern.ch/previous-schools}}
\rfoot{\thepage\hspace*{3mm}}
 
}

\usepackage{dcolumn}
\newcolumntype{d}{D{.}{.}{2.5}}
\newcolumntype{s}{D{.}{.}{1.2}}

\begin{document}

\title{Beam Cleaning and Collimation Systems}

\author{S. Redaelli}

\institute{CERN, Geneva, Switzerland}

\begin{abstract}

Collimation systems in particle accelerators are designed to safely and efficiently dispose of unavoidable beam losses during operation. Their specific roles vary depending on the type of accelerator. The state of the art in hadron beam collimation for high-intensity, high-energy superconducting colliders is exemplified by the system implemented at the CERN Large Hadron Collider (LHC). In this machine, the stored beam energy reaches levels several orders of magnitude higher than the tiny energy required to quench superconducting magnets. It also exceeds by orders of magnitude the damage thresholds of typical accelerator components, placing stringent demands on beam loss control.
Collimation systems are therefore essential for the reliable daily operation of modern accelerators. This lecture reviews the design of a multistage collimation system, using the LHC as a case study. The LHC collimation system has achieved unprecedented cleaning performance, with a level of complexity unmatched by any other accelerator. Design aspects and operational challenges of such large-scale collimation systems are also discussed.
\end{abstract}


\maketitle
\thispagestyle{ARTTITLE}


\section{Introduction}
\label{intro}

\begin{figure}[h]
  \centering
  \includegraphics[width=0.65\linewidth]{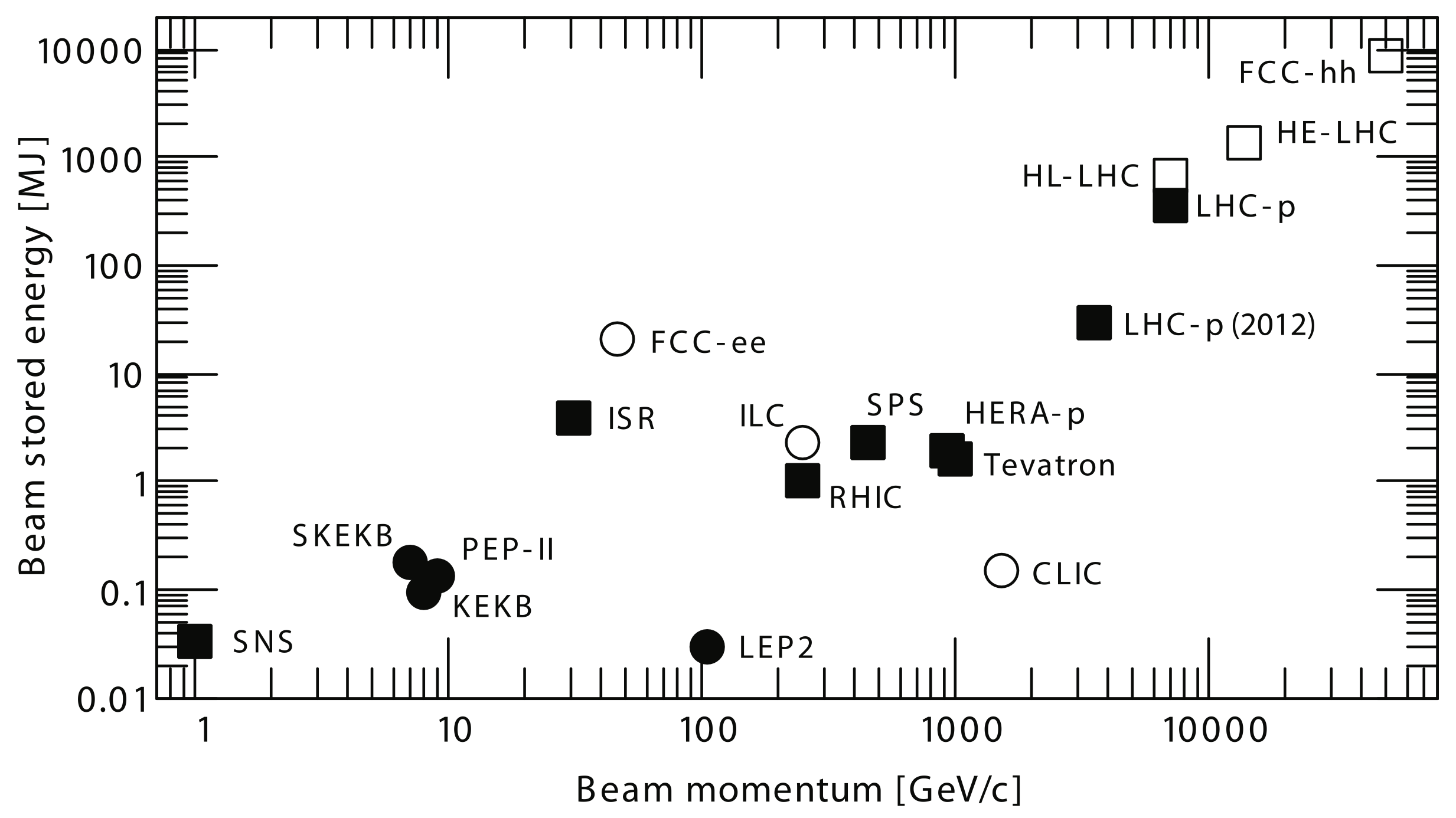}
  \vspace{-0.2cm}
  \caption{Livingston-like plot of beam stored energy for hadron (squares) and lepton (circles) particle accelerators. Filled symbols are used for past or operating machines and empty symbols indicate future accelerators. {\it Adapted from an initial version by R.~A\ss mann.}}
  \vspace{-0.2cm}
  \label{fig_eb}
\end{figure}

The importance of beam collimation systems in modern particle accelerators has grown significantly in the pursuit of higher beam energies and intensities. For reference, the stored beam energy of recent and future accelerators is illustrated in~\Fref{fig_eb}. The design value for the CERN Large Hadron Collider (LHC)~\cite{lhc} is 362~MJ. This target has already been surpassed, with the LHC reaching 430~MJ during proton operation at 6.8~TeV in 2023. The goal for the High-Luminosity LHC (HL-LHC) upgrade is about $700~\mathrm{MJ}$—nearly twice the LHC original design value~\cite{hl}.
The HL-LHC therefore defines a new frontier for beam collimation requirements~\cite{Redaelli:2020mld}. Accelerators with such large stored beam energies, particularly supercolliders~\cite{Shiltsev:2019rfl} based on superconducting magnet technology, simply cannot operate without efficient collimation systems to control the unavoidable losses that occur during normal beam operation.

The operation and physics goals of modern superconducting, high-energy hadron colliders—such as the Tevatron \cite{tev}, the Relativistic Heavy Ion Collider \cite{rhic}, and the LHC—could not be achieved without adequate beam collimation. With the LHC and its upcoming HL-LHC upgrade, both the design complexity and the performance requirements of beam collimation have reached unprecedented levels. Efficient cleaning of beam losses is essential to prevent particles from reaching the small apertures of superconducting magnets.
An effective collimation system is also crucial for maintaining high machine availability, as it minimizes downtime caused by beam losses that are not properly intercepted or controlled. The work carried out so far, together with the operational experience gained from these machines, provides a solid foundation for the design of collimation systems in future colliders.

\begin{figure}
  \centering
  \includegraphics[width=95Mm]{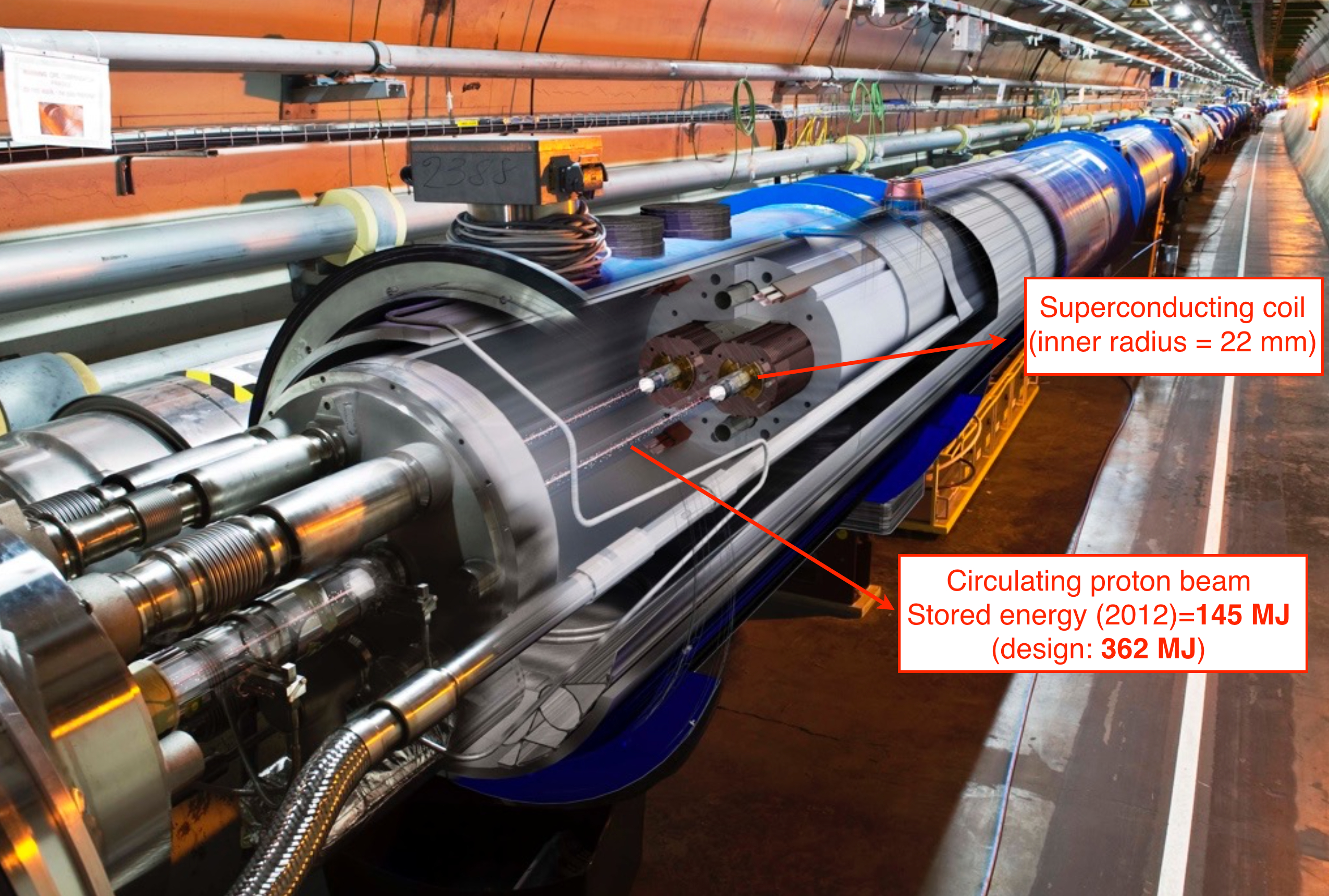}
  \vspace{-0.2cm}
  \caption{The LHC dipole in the tunnel, showing the cross-section of the magnet cold mass }
  \label{fig_dip}
\end{figure}

To illustrate the challenges of beam collimation, \Fref{fig_dip} shows a schematic view of an LHC superconducting dipole magnet in the underground tunnel. The 3D cutaway highlights the main internal components, in particular the superconducting coil that surrounds the beam vacuum chamber. The inner radius of the chamber through which the beam travels is about 17~mm vertically and 22~mm horizontally. This means that the inner aperture of the LHC magnets is only about 2~cm away from a circulating beam storing an energy more than a billion times greater than the small energy deposition needed to disturb superconducting operation. The quench limit of these magnets is estimated to be only 20–30~mW/cm$^{3}$—to be compared with hundreds of megajoules of stored beam energy. This striking contrast makes it clear that transverse beam losses must be extremely well controlled. The population of particles in the beam halo must decrease rapidly with increasing transverse amplitude, ensuring that very few particles stray far from the nominal beam orbit at the centre of the pipe.

This document, which updates a previous lecture~\cite{Redaelli:2016ohw}, presents the design principles of collimation systems for hadron accelerators, with particular emphasis on the requirements and design aspects of high-energy and high-intensity machines. The primary purpose of a collimation system is to safely and systematically dispose of beam losses that would otherwise occur at sensitive locations or on equipment not designed to withstand direct particle impact. In practice, this general concept is implemented differently depending on the specific performance requirements of each accelerator. For example, collimation in warm high-power machines focuses on localizing beam losses within designated regions to prevent widespread radiation and protect surrounding equipment. In contrast, superconducting accelerators require losses in cold magnets to remain below quench limits to ensure stable operation. In colliders, additional requirements arise from the need to minimize beam-loss-induced background in the~experimental detectors.

The design goals of collimation systems in accelerators are discussed in Section~\ref{roles}. In Section~\ref{notation}, the required notation is introduced and the inputs to collimation design from machine aperture and beam loss mechanisms are discussed: aperture model, loss assumptions and definition of performance targets are presented. The design of a multistage collimation system is outlined in Section~\ref{stages}. In the following, the document is focused on the presentation of the LHC collimation system as a case study. In Section~\ref{lhc-coll}, the system layout is reviewed and operational challenges for beam collimation are introduced, presenting the solutions deployed. The LHC collimator design is discussed in Section~\ref{design} and the collimation performance achieved in the LHC is reviewed in Section~\ref{lhc-perf}. Conclusive remarks are drawn in Section~\ref{concl}.


Although not discussed in detail in this lecture, collimation is becoming increasingly important for lepton accelerators as well. While their stored beam energies are typically several factors lower than those of hadron machines (see~\Fref{fig_eb}), existing facilities still face significant challenges in managing beam losses. Recent examples include SuperKEKB~\cite{Terui:2024rdo}. For future projects such as the electron–positron Future Circular Collider (FCC-ee)~\cite{Andre:2025bpv}, dedicated studies are underway to develop suitable beam collimation solutions~\cite{Abramov:2023hmw}. In these accelerator, an additional concern arises from the need to control backgrounds affecting the experiments. Integrated approaches that combine collimation and background mitigation are therefore being explored. The concepts and design strategies presented in this lecture for high-performance multistage collimation systems are also applicable to lepton colliders and are already being used to optimize the collimation design of future facilities.


\section{Design goals for collimation systems in particle accelerators}
\label{roles}

The typical design goals of collimation systems can be summarized as follows. As mentioned above, specific collimation solutions must be studied independently for each accelerator. Performance requirements are usually defined by identifying the relevant beam loss scenarios and the most critical requirements that the system must be designed for. For example, in the LHC, the primary design goal of the initial collimation system was to allow efficient operation without quenching the superconducting magnets. Other important functions, such as protecting equipment and minimizing backgrounds, were naturally achieved through the optimized design developed to meet this requirement. In other machines, such as the Tevatron, the main focus of the collimation system was instead on reducing experimental backgrounds.

\begin{itemize}
\item \textbf{Cleaning of betatron and off-momentum beam halos}: Unavoidable beam
  losses of halo par\-ticles must be intercepted and safely disposed of before
  they reach sensitive equipment. The~required cleaning performance
  depends on the design of the accelerator. The most challenging require\-ments
  arise for superconducting accelerators, where loads from
  beam losses must remain  below the quench limits of superconducting magnets.
  For
  example, the LHC design beam stored energy in the range of hundreds of MJ has to be compared with
  typical quench limits of a few tens of mW/cm$^3$ \cite{q-paper}. See \Fref{fig_dip} for a graphical view of the LHC challenge. Similar challenges apply to the RHIC and the Tevatron despite of the lower stored beam energy values. For accelerators based on normal conducting magnets, concerns are mainly related to activation of components. 
\item \textbf{Passive machine protection}: Collimators are the closest elements
 to the circulating beam and repre\-sent the first line of defense in various normal and abnormal loss cases, scenarios. Owing to the damage potential of high-intensity beams, this
 functionality has become one of the most critical aspects of the operation
 of accelerators \cite{rs, jw}, as well as a crucial input to the design of
 collimators that must withstand design failures. Robust collimators must intercept safely beam losses in case of failures, keeping energy that leaks to other equipment below their damage limits.
\item \textbf{Cleaning of collision products}: In colliders, this is achieved
  with dedicated fixed and movable collimators located in the outgoing beam paths of each
   experiment, to catch the products of collisions: direct
  collision debris and beam particles that emerge from the collision points
  with modified angles and energy. The main goal in this case is controlling the irradiation of components, as the~luminosity performance shall not be compromised: losses from the collision debris are unavoidable and are part of a collider operation.
\item \textbf{Optimization of the experiment background} (\ie minimization of
  halo-induced noise in de\-tector measurements): this is one of the key target performance figure
  for collimation systems in previous col\-liders, like ISR~\cite{Hubner:2012td}, SppS~\cite{Schmidt:2016ivs} and Tevatron.
  Efficient suppression of tails or local shielding at the detector locations
  can reduce spurious signals in detectors (see, for example, \cite{bckg}). 
\item \textbf{Concentration of radiation doses}: for high-power machines, it
  is becoming increasingly import\-ant to be able to localize beam losses in
  confined and optimized `hot' areas rather than having a~distribution of
  many activated areas along the machine. Fulfilling this requirement allows easy access for maintenance in the largest fraction of the machine.
\item \textbf{Local protection of equipment for improved lifetime against radiation
  effects}: Dedicated mov\-able or fixed collimators are used to shield
  equipment locally. For example, passive absorbers are used in the LHC collimation
  inserts to reduce total doses, and to warm dipoles and quadrupoles
  that would otherwise have a short lifetime in the high-radiation environment
  foreseen during the~nominal LHC operation. The exposure of radiation to equipment might not
  pose immediate limi\-tations to operation of a machine but its optimization is crucial
  to ensure long-term reliability.
\item \textbf{Beam halo scraping and halo diagnostics}:
  Although rarely a primary design criterion, the ability to actively scan the beam distribution can be a very useful feature of a collimation system. When combined with sensitive beam loss monitoring, collimator scanning provides a powerful method to probe the population of beam tails~\cite{diff, mess}, which are otherwise too small relative to the beam core to be measured using conventional emittance measurement techniques. Thanks to their robustness, collimators can also be used efficiently to scrape the beam for tail shaping and intensity control. For example, this technique is applied before extraction from the SPS to fine-tune the beam prior to injection into the LHC. Another example of collimator scraping for background control at the~LHC can be found in~\cite{2s,Mirarchi:2020qjb}.
  
\end{itemize}

A well-designed collimation system can often fulfill several roles simultaneously. For example, concentrating radiation losses or reducing experimental backgrounds are naturally achieved by highly efficient betatron and off-momentum collimation systems, which minimize leakage to other areas of the~machine, particularly the experimental regions. It is noteworthy that the present LHC beam collimation system~\cite{lhc, finalColl} is unique in fulfilling all the roles above, thanks to a careful design that extends beyond simple cleaning functionality. The cost of this high performance is unprecedented system complexity, which poses significant operational challenges, as discussed in Section~\ref{lhc-perf}.

This lecture focuses on beam collimation systems and does not explicitly cover injection or dump protection systems. These systems also employ beam-intercepting devices, which safely absorb beam losses in the event of fast failures during beam injection or extraction, typically caused by malfunctions of the fast kicker magnets~\cite{aserinjection2019}. Injection and dump protection are closely related to the collimation system because they are designed to protect the same aperture bottlenecks, and their settings must be coordinated with those of the collimation system; all these devices are arranged with respect to the circulating beam orbit. However, specific design criteria apply. In particular, injection and dump protection must withstand the largest single-pass beam loads, requiring especially robust designs. The interplay between these systems is further discussed in Section~\ref{sec:multistage}.
\newpage
\section{Inputs to collimation design from beam loss mechanisms and machine aperture}
 \label{notation}


\subsection{Basic definitions for collimation and beam halo}
\begin{figure}[!b]
  \centering
  \includegraphics[width=80mm]{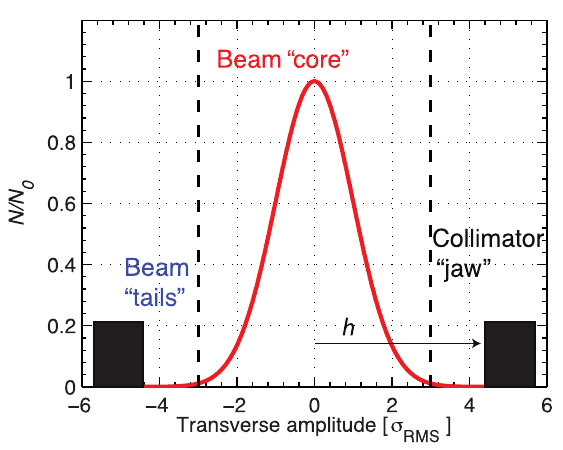}
  \vspace{-0.2cm}
  \caption{Normalized Gaussian distribution often 
    used to model the beam-core particle distribution (red line).
    Overpopulated tails may be intercepted by collimator jaws, which
    constrain particle motion at a given amplitude.}
  \label{fig_gau}
\end{figure}

Particles with transverse amplitudes or energy deviations significantly larger
than those of the reference particle are referred to as \emph{beam halo particles}.
One can distinguish between \emph{betatron} and \emph{off-momentum} halos, which are formed in
the case of larger-than-nominal transverse emittance or energy errors, respect\-ively. In practice, both components are typically present and one can hardly identify a clear cut between these two halo components.
The transverse amplitude of a particle $i$ around the reference closed orbit,
$z\equiv(x,y)$, can be expressed as a function of the longitudinal curvilinear
coordinate $s$ for the Twiss parameters $\beta_z(s)$, $D_z(s)$ , and $\phi_z(s)$
as
\begin{equation}
z_i(s) = \sqrt{\beta_z(s)\epsilon_{z, i}}\sin[\phi_z(s)+\phi_{z,i,0}] +
\left(\frac{\delta p}{p}\right)_iD_z(s)~,
\label{eqMot}
\end{equation}
where $\epsilon_{z, i}$ is the single-particle emittance,
$\left(\delta p/p\right)_i$ is the energy error, and $\phi_{z,i,0}$ is
an arbitrary phase. The r.m.s.~size of the beam at location $s$ is then
given by
\begin{equation}
\sqrt{\beta_z(s)\epsilon_z+\left(\frac{\delta p}{p}\right)^2D_z^2(s)}~,
\end{equation}
where $\epsilon_z$ and $\delta p/p$ are the r.m.s.~transverse emittance
and energy spread of the beam. 
Machine apertures and collimator settings are expressed, unless specified otherwise, units of \emph{betatron
  beam size},
\begin{equation}
  \sigma_z(s)=\sqrt{\beta_z(s)\epsilon_z}~,
  \label{sz}
\end{equation}
which takes into account only the contribution to the beam size from the
betatron motion. Collimator settings might then be given in normalized units
as
\begin{equation}
  n_\sigma=\frac{h}{\sigma_z}~,
  \label{ns}
\end{equation}
where $h$ is the distance in millimetres between the collimator jaw and the circulating
beam, \eg the half gap of a two-sided collimator centred around the beam orbit, as
shown in \Fref{fig_gau}. The calculation of $n_\sigma$ settings for ``skew'' collimators that are not purely horizontal or vertical is discussed in Section~\ref{sec:coll-sett}.

The distinction between halo and core particles is, to a certain extent,
arbitrary. For Gaussian distributions, one may define as halo particles those
with amplitudes above three r.m.s.~deviations of the Gaussian (see dashed
lines in \Fref{fig_gau}), \ie with emittances larger than
$9\epsilon_z$.
For a beam with perfect two-dimensional Gaussian distributions in the $(z,z')$ plane,
about 1.1\% of the total beam particles have amplitudes above $3\sigma_z$
and 0.03\% above $4\sigma_z$, respectively. Particles this far out from the beam
core are rarely of any practical use for the accelerator and are more likely to cause concerns (beam
losses, irradiation of components, background in detectors, \etc). In colliders, their contribution to luminosity is regarded here as negligible, although not null, and potentially detrimental.

For off-momentum halos, a similar definition could be adopted to express the distance of halo particle from the beam core in terms of the r.m.s~distribution of the beam's energy spread. For most practical 
purposes in circular accelerators, one might consider as off-momentum halo the particles
outside the RF bucket that are lost when beams are
accelerated or in the presence of synchrotron radiation (which is non-negligible at the
LHC). 

Beam collimation is achieved by placing blocks of material, the \emph{collimator
  jaws}, close to the circu\-lating beams, to constrain the amplitudes of
stray particles outside the core. This is shown schematically by the black
boxes in \Fref{fig_gau}. Collimation of off-momentum tails might be
achieved in a similar way as for betatron tails, by placing collimators at
locations of high dispersion, where the particle's energy shift results
in a transverse offset, as in the second term on the right-hand side of
\Eref{eqMot}. Special optics conditions must be respected to ensure that off-momentum halos can be collimated efficiently while respecting the betatron cut and ensuring that no ``core beam particles'' are intercepted in the off-momentum collimation process. This is typically achieved by maximising the normalized dispersion, $\frac{D_x(s)}{\sqrt\beta_x(s)}$, at the off-momentum collimation system location.


How close a collimator should be to the beam depends on several factors, which will be discussed in detail later in this paper. The outer limit of a collimator setting is determined by the available machine aperture that must be protected and by the level of cleaning performance required. The machine aperture is therefore one of the key design inputs for the beam collimation system. The inner limit of collimator settings is mainly set by how the collimators affect beam stability through an increase in machine impedance. Tighter settings also tend to produce higher beam losses and impose stricter tolerances for orbit and optics errors. As a general guideline, collimators should be positioned no closer to the beam than strictly necessary and should avoid intercepting particles that are still useful for the accelerator, for example to produce luminosity in a collider. Aspects related to collimation impedance are not covered in detail here, but can be found in companion lectures at this school.

\subsection{Collimation performance definition: cleaning inefficiency}
The cleaning performance of a collimation system is measured by the
\emph{collimation efficiency}, a figure of merit that expresses the fraction
of halo particles caught by the system over the total lost from the beam.
A perfect beam collimation provides 100\% cleaning efficiency, \ie there is no beam loss
at sensitive equipment for any circulating-beam losses (also referred to as {\it primary beam losses}). Such a system, of course, does not exist. So, in the collimation design phase one needs to define an adequate target for the cleaning-efficiency performance by understanding what are tolerable losses at different exposed elements around the ring. This depends also on loss assumptions, as discussed in the next sections.

In order to study collimation losses at different locations of the ring, the
\emph{cleaning inefficiency}, $\eta_\text{c}$, can be introduced as the relative
fraction of beam that leaks to other accelerator components, $A_{\text{lost}}$,
compared with what is intercepted and safely disposed of by the
collimators, $A_{\text{coll}}$:
\begin{equation}
  \eta_\text{c}=\frac{A_{\text{lost}}}{A_{\text{coll}}}~.
\end{equation}
The relevant measure of beam loss, indicated by the letter $A$ in this equation,
has to be identified for the~specific design criteria that the collimation
system addresses -- quench limit of superconducting magnets, background to experiments or activation of components just to give some examples.

The LHC beam collimation requirements are driven by the challenge to keep beam
losses below the quench limits of the superconducting magnets, \ie the figure of merit $A_{\text{lost}}$ is the energy deposited in the superconducting coils of exposed magnets. For a semi-analytical treatment that is used in early design phases, it is useful to express quench limits and collimation inefficiency in terms of protons lost per metre. In this case, the
inefficiency $\eta_\text{c}$ is then defined as the number of protons lost as a fraction of
the total number of particles absorbed by the collimation system. The {\it
  local cleaning inefficiency}, $\tilde\eta_\text{c}\equiv\tilde\eta_\text{c}(s)$, is defined
as a function of the longitudinal coordinate $s$ as the fractional loss per unit
length,
\begin{equation}
  \tilde\eta_\text{c}=
  \frac{N(s\rightarrow s+\Delta s)}{N_{\text{abs}}}\frac{1}{\Delta s}~,
  \label{eta}
\end{equation}
where $N(s\rightarrow s+\Delta s)$ is the number of particles lost over the distance
$\Delta s$, \ie in the longitudinal range $(s, s+\Delta s)$, and $N_{\text{abs}}$
is the number of particles absorbed by the collimation system. 
The longitudinal distribution has to be evaluated over an adequate {\it dilution length} depending on various layout considerations, \eg over the length of a magnet.

This definition has the advantage that it can be directly compared against the quench
limits of super\-conducting magnets if a proper \emph{dilution length} is
chosen. Indeed, for the LHC it was estimated \cite{RalphCham12} that
the quench limits in units of proton lost per metre, $R_\text{q}$, are
\begin{eqnarray}
R_\text{q}^{\text{inj}}&=&7.0\times10^{8}\:{\text{protons}/(\UmZ\cdot\UsZ)}\:\:{(450\UGeV)}~, \\
R_\text{q}^{\text{top}}&=&7.6\times10^{6}\:{\text{protons}/(\UmZ\cdot\UsZ)}\:\:{(7\UTeV)}~,
\label{q}
\end{eqnarray}
for beams at injection (${\text{inj}}$) and top (${\text{top}}$) energies, respectively. These
approximate figures were used in the early LHC design phase and
in first collimation performance estimates \cite{cham2005}.
Although nowadays detailed simulation tools and more precise models are
available to evaluate the energy deposition in the~magnet coils for a 
direct comparison against quench limits of superconducting cables (see, for example, Ref.~\cite{q-paper}), the formalism introduced here is very useful and pedagogical. 


\subsection{Beam loss modelling through beam lifetime}

The performance specifications of a collimation system start with establishing, under well-defined assumptions, a beam loss model for the accelerator. The potential damage to components or the quenching of superconducting magnets must be evaluated against the local power deposition. For a given collimation cleaning performance, this depends on the total primary-beam losses, which are estimated through the construction of a so-called beam loss model.

The design LHC beams, storing 362~MJ, correspond to about 77~kg of TNT and could melt nearly 500~kg of copper~\cite{tnt}. The beam loss model must also take into account the duration of losses, since no material or component can withstand the full LHC beam if deposited over short time scales. The~impact of beam losses depends critically on their temporal profile. Establishing design loss criteria for new accelerators is therefore challenging, and conservative assumptions are typically adopted during the early design phases to ensure adequate protection margins against both regular losses and failure scenarios. It is worth noting that the design goal is not to dispose of the entire LHC beam, but only a fraction of the total stored energy as determined by the beam loss model. This fraction is calculated below. If the actual loss rates exceed the design limits, an emergency beam extraction is triggered. At the LHC, several systems redundantly monitor these conditions~\cite{rs, jw}, including beam loss monitors distributed around the ring and beam current measurement systems.

There are various mechanisms that can lead to beam losses in particle accelerators~\cite{rs, jw}. A~useful distinction can be made between \emph{regular} and \emph{abnormal} losses: the former refer to the unavoidable losses that occur during standard operation, while the latter arise from failures of accelerator systems or from incorrect beam manipulations. Losses in both categories can occur over a wide range of timescales, from a fraction of a single revolution to several tens of seconds.

In circular colliders, a primary source of beam losses originates from collisions between the opposing beams, which cause the \emph{burn-off} of beam particles. Other loss mechanisms include interactions with residual gas, intra-beam scattering, and various types of beam instabilities (single-bunch, collective, beam–beam, etc.). Additional contributions arise from noise in the feedback systems used to stabilize the beams, as well as from transverse and longitudinal resonances, including RF noise.

Further losses occur during normal operational phases of the accelerator—such as particle capture at the beginning of the ramp, injection and beam dump processes, or dynamic changes in the operational cycle (orbit drifts, optics adjustments, energy ramps, etc.). These are collectively referred to as \emph{operational losses}~\cite{vk}.
Examples of abnormal losses include those triggered by magnet failures while the~beams are circulating. Their characteristic timescales vary significantly depending on machine parameters. Very fast abnormal losses can also be induced by the onset of beam instabilities or by failures of injection and dump kicker magnets. The latter can cause \emph{single-turn losses} if the bunched beam encounters the rising or falling edge of a kicker pulse. For the LHC, the most critical failure scenario is an~asynchronous firing—that is, a firing of the dump kickers not synchronized with the abort gap. If such a failure occurs during the passage of a bunch train, the rising field can deflect several bunches to large amplitudes. These bunches then complete one full turn around the ring before encountering the~correct kicker field at the dump. Since individual LHC bunches at 7~TeV exceed the damage threshold of metals by far, special collimators are installed to mitigate the consequences of such events. Moreover, the~machine configuration for high-intensity operation must include sufficient aperture margins to prevent permanent damage to any component (this topic is discussed in detail in~\cite{roderikBeta}).

\begin{figure}
  \centering
  \includegraphics[width=65mm]{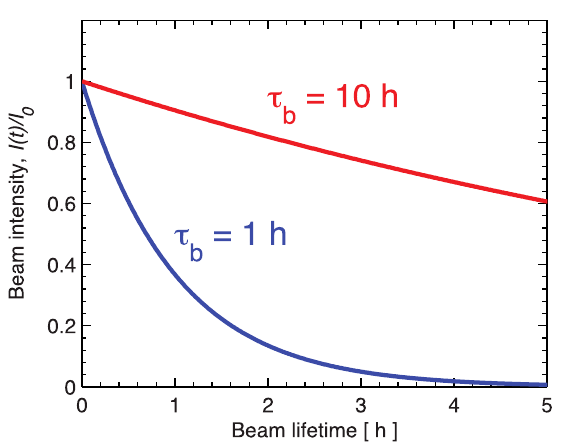}
  \vspace{-0.2cm}
  \caption{Relative reduction of beam current versus time, $I(t)/I_0$, for
  beam lifetime values of $1\Uh$ and $10\Uh$}
  \label{fig_lt}
\end{figure}

Ignoring, for the moment, very fast loss scenarios and their
impact on collimator design \cite{rs, jw}, let us consider 
beam collimation in the presence
of \emph{diffusive losses}. In this case, the~transverse increase
of particle amplitude per turn is much smaller than, say, one sigma of the~r.m.s.~distribution. Rather than treating each loss mechanism in detail, losses are modelled by considering the {\it beam lifetime}.
The~time-dependent circulating beam intensity, $I(t)$, can, for most practical
purposes, be modelled by an~exponential decay function whose time constant,
$\tau_\text{b}\equiv\tau_\text{b}(t)$, defines the~\emph{beam lifetime} as
\begin{equation}
I(t) = I_0\mathrm{e}^{-t/\tau_\text{b}},
\end{equation}
where $I_0$ denotes the initial beam current. After a time $\tau_\text{b}$, the total beam
current is reduced to about 37\%. Example profiles of relative beam intensity versus time, $I(t)/I_0$,  are shown in \Fref{fig_lt} for lifetime
values of $1\Uh$ and $10\Uh$. In a linear approximation, beam loss rates, $\text{d}I/\text{d}t$,
are inversely proportional to $\tau_\text{b}$ and can be calculated as
\begin{equation}
  -\frac{1}{I}\frac{\text{d}I}{\text{d}t} = \frac{1}{\tau_\text{b}}~.
\end{equation}
It is important to emphasize that the beam lifetime, $\tau_\text{b}(t)$, is indeed a function of time
and is not constant through the oper\-ational cycle. The sources of beam losses
introduced previously -- operational losses and other acceler\-ator physics
mechanisms -- occur at
different times in the operational cycle and might become apparent as drops of beam lifetime at given
times in the cycle. 

\begin{figure}
  \centering
  \includegraphics[width=110mm]{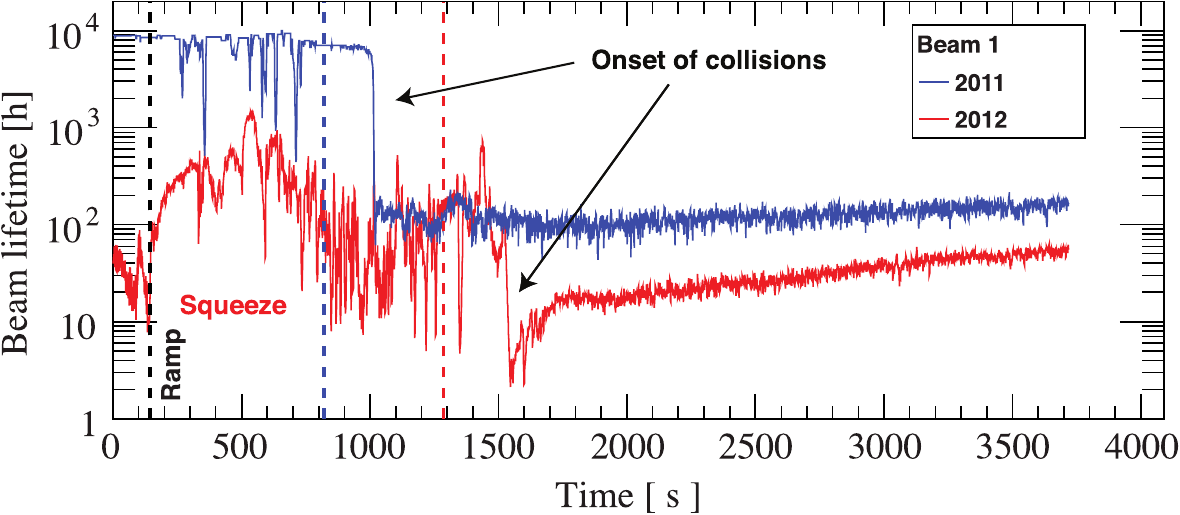}
  \includegraphics[width=110mm]{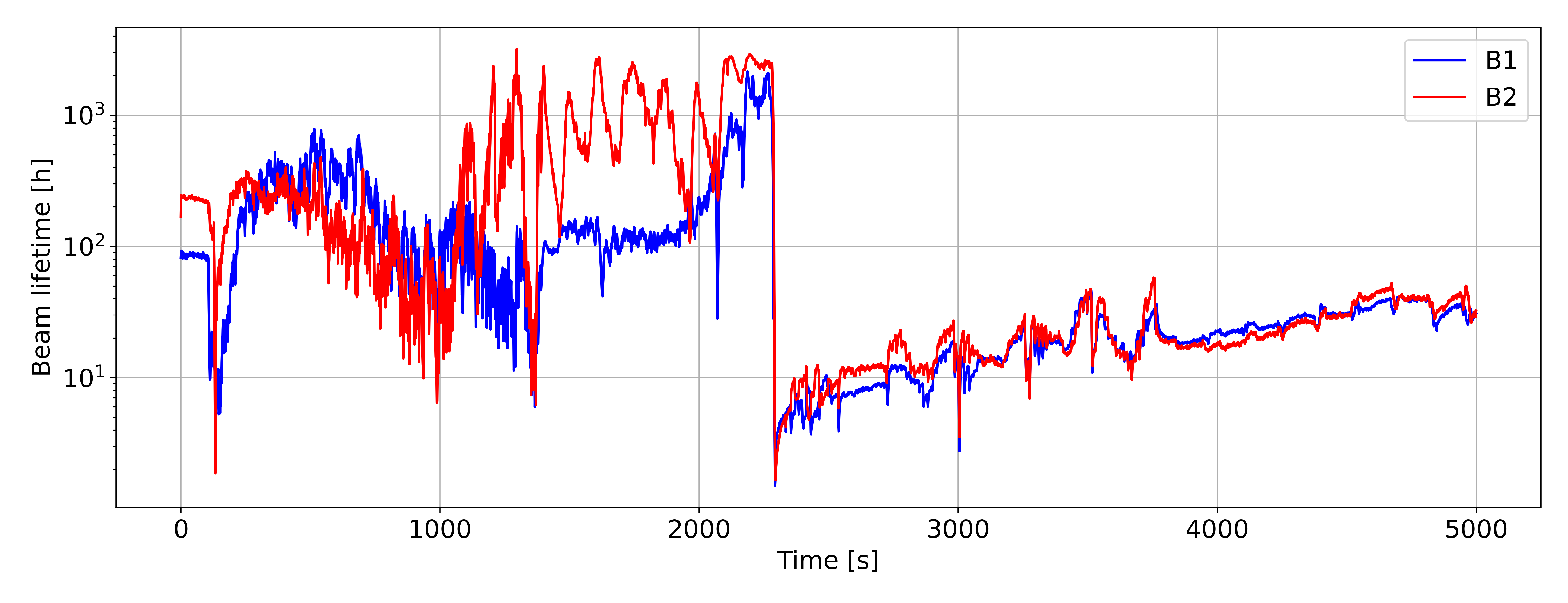}
  \vspace{-0.2cm}
  \caption{Measured beam lifetime as a function of time during two typical LHC physics fills in 2011 at $3.5\UTeV$ (top graph, blue) and in 2012 to $4\UTeV$ (top graph, red) for beam~1 and in a fill in 2025 for both beams (bottom graph). {\it Courtesy of B.~Salvachua}~\cite{Salvachua:2013uua} and {\it S.~Morales~Vigo}.
}
  \label{fig_lt_meas}
\end{figure}

An example of the measured beam lifetime, $\tau_\text{b}$, during LHC physics fills is shown in~\Fref{fig_lt_meas}, comparing operational runs from 2011 and 2012. A representative example from 2025 is also included. Similar qualitative features are observed in other operational years. The different machine modes are indicated in the top graph. In this example, a characteristic ``spiky'' behaviour of the lifetime is seen, with varying values during the energy ramp and squeeze, while $\tau_\text{b}$ remains well above a few hours. The smallest values typically occur at the onset of collisions, when the transverse beam separation is collapsed to bring the opposing beams into full overlap. After this transient phase, a smooth evolution of $\tau_\text{b}$ is generally observed.
The LHC operational cycle has increased in complexity in recent years. During the ongoing Run~3 (2022–2026), the machine has reached its most intricate configuration so far, featuring simultaneous luminosity levelling based on optics, transverse separation, and crossing angle adjustments. This results in a less smooth time profile than in previous years, which had simpler operational modes. Nevertheless, the beam lifetime remains consistently well above a few hours.

A collimation system must be designed to cope with the maximum expected rates of
beam loss. This is determined by the \emph{minimum allowed beam lifetime},
$\tau_\text{b}^{\min}$, throughout the operational cycle, most not\-ably
during phases at maximum energy (flat-top, squeeze, collision
preparation, and physics data recording) when the beam stored energy is the largest.
The design value used to specify the LHC collimation system is
$\tau_\text{b}^{\min}=0.2\Uh$ for up to a maximum time of $10\Us$ \cite{RalphCham12}.
Values of $\tau_\text{b}$ below $1\Uh$ were recorded on a regular
basis during the LHC operation, on the other hand the recent operational experience show that this assumption is conservative for the estimation of beam losses. On the other hand, it is essential to remain conservative in early phases of the collimation design given the uncertainties in extrapolating the operational experience to more-pushed configurations, like the new HL-LHC beam conditions expected for the LHC upgrade.



Let us calculate the peak design loss that the LHC collimation system must withstand when the~minimum allowed beam lifetime is reached at full intensity and 7~TeV. The LHC design~\cite{lhc} foresaw ramping 2808 bunches with $1.15\times10^{11}$ protons per bunch, corresponding to a stored beam energy of 362~MJ, or an equivalent beam power of 4065~GW. For a beam lifetime of 0.2~h, primary beam losses reach a peak level of $I_0/\tau_b = 4.5\times10^{11}$ protons lost per second, corresponding to 502~kW. This loss level must be sustainable without causing quenches or permanent damage to any machine component, including the collimators, for up to 10~s. It is assumed that the beams would be dumped if such a high-loss regime were maintained for longer durations than this specification, without the possibility of recovering an acceptable lifetime. If the beam intensity were constant, the total beam power loss over 10~s would amount to approximately 5.0~MJ. In practice, the correct power loss is computed by integrating the curves in~\Fref{fig_lt}, although the correction is below the 1~\% level for the time scale considered here.

It will be shown below how these numbers, in particular the estimated peak loss rate at maximum beam current, are used to assess the quench performance of an accelerator like the LHC for a given cleaning performance of the collimation system.




\subsection{Machine aperture model and impact on collimation settings}
\label{Sec:apert}
In particle accelerators, an {\it aperture model} is needed to define beam clearance in each vacuum beam pipe of every accelerator component. In high-energy accelerators, the aperture model is of utmost importance to provide quantitatively the transverse space available for the reliable operation of the circulating beams while identifying adequate margins for the stable and safe operation. The physical dimensions of the~mechanical apertures are of course the primary input to establish an accurate aperture model. The~beam and optics properties, other relevant beam dynamics factors like betatron oscillations, dispersion, and nonlinear effects, as well as realistic tolerances (manufacturing tolerances, alignment errors, optics errors and beam dynamics perturbations) have as well to be accounted for. 

It is important to underline that the aperture model is essential for predicting beam losses, optimizing machine performance, and ensuring safe operation, particularly in high-intensity accelerators where beam halo and instabilities can cause significant damage or radiation. For the LHC, the so-called $n_1$ formalism was developed to take into account consistently all the required inputs to the aperture model~\cite{n1}. For convenience, in most cases the aperture is defined in units of betatron beam size as defined in Eq.~(\ref{sz}). Using this {\it normalized} aperture is a well-established approach that has the advantage to avoid dealing with different numbers in millimeters for the varying mechanical apertures and beam dimensions around the ring. While detailed aperture calculations are beyond the scope of this lecture, some salient results are shown without derivation. 

The primary input from the aperture model to the collimation design is the minimum aperture to be protected, estimated in every phase of the accelerator operational cycle. This is often referred to as {\it the~aperture bottleneck} of the accelerator. The collimation system must ensure that no aperture bottleneck is exposed to primary beam losses and that the local cleaning inefficiency at every ring location, as defined in Eq.~(\ref{eta}), is adequate and compatible with the operation at high intensity. Below we will demonstrate that this is only possible with a multi-stage collimation system composed, for example, of primary (TCP), secondary (TCS) and tertiary (TCT) collimators. If $n_{\rm BTNK}$ is the normalized aperture at the~bottleneck, the aperture protection with required cleaning performance is ensured only if the~condition
\begin{equation}
    n_{\rm TCP}<n_{\rm TCS}<n_{\rm TCT}<n_{\rm BTNK}
    \label{EqHierarchy}
\end{equation}
is respected\footnote{This statement assumes that the population of the tertiary beam halo, leaking }. Here, subscripts indicate the {\it collimation families} introduced above. Respecting the so-called {\it collimation hierarchy} of Eq.~(\ref{EqHierarchy}) -- which is some cases even includes more collimation stages -- is one of the main operational challenges for accelerators like the LHC.

Aperture calculations carried out in the early LHC design phase are reviewed here to put the discussion in a quantitative context. The minimum apertures calculated~\cite{cham2005} for both planes and beams at 450~GeV (injection) and 7~TeV (top energy) are listed in Table~\ref{tab_bottleneck_inj}. The simulations were performed for both warm and cold elements, as the local cleaning requirements are very different for these two cases: the former can obviously tolerate larger losses, leading to different specifications for beam collimation. The calculations relied on error models for the $n_1$ formalism~\cite{n1}, which eventually turned out to be conservative, as demonstrated by later beam-based aperture measurements~\cite{ap2, ap3}. As already mentioned for the lifetime assumptions, it is important to remain conservative in the early design phases.

In the LHC, according to the $n_1$ aperture model, the aperture bottleneck values are similar at injection and at top energy under collision conditions. Estimated values range from 6.7~$\sigma_z$ to 8.9~$\sigma_z$. However, the locations of the bottlenecks are very different. At injection, the model predicts distributed bottlenecks around the arcs, limited by the apertures of superconducting dipoles and quadrupoles, while at top energy the bottlenecks occur in the inner triplets of the high-luminosity experiments~\cite{lhc}. Here, the beta functions at the aperture bottlenecks — the superconducting inner triplet magnets — are brought to values above 1~km in the so-called {\it squeeze} process in order to produce small beams at the collision points.

The key observation from the calculations in Table~\ref{tab_bottleneck_inj} is that the aperture to be protected is, by design, quite small. This is, of course, the result of a cost-optimization exercise in the design phase of the superconducting magnets, in particular for the arc quadrupoles (308 magnets) and dipoles (1232 magnets). The small estimated aperture poses tight challenges for the collimation hierarchy, as will be discussed later: in order to satisfy the conditions of Eq.~(\ref{EqHierarchy}), typical settings of the primary collimators are 5.0~$\sigma_z$ to 6.0~$\sigma_z$. As an important side note, since the optics in the collimator insertions are typically not changed during the operational cycle, the shrinkage of the beam emittance during the energy ramp demands moving the collimators to maintain nearly constant normalized settings.

\begin{table}
  \caption{Minimum horizontal and vertical apertures at injection
  (450\UGeV) and top energy (7\UTeV, $\beta^*=0.55\Um$) for warm and cold
  elements, as estimated in the LHC design phase \cite{cham2005}. }

  \begin{center}
    \begin{tabular}{lddcdd}
       \hline\hline
      & \multicolumn{2}{c}{\textbf{450\,GeV}}&&\multicolumn{2}{c}{\textbf{7\,TeV}}\\
     \cline{2-3}\cline{5-6}
      &  \multicolumn{1}{l}{\textbf{Warm}}  &  \multicolumn{1}{l}{\textbf{Cold}} & &  \multicolumn{1}{l}{\textbf{Warm}}  &  \multicolumn{1}{l}{\textbf{Cold}}  \\
      \hline
     \textbf{Beam 1}\\
      Horizontal   & 6.8 & 7.9 && 28 & 8.9\\
      Vertical     & 7.7 & 7.8 && 8.3 & 8.4\\

      \textbf{Beam 2}\\
      Horizontal   & 6.7 & 7.7 && 28 & 8.1\\
      Vertical     & 7.7 & 7.6 && 8.7 & 8.8\\
      \hline\hline
    \end{tabular}
  \end{center}

  \label{tab_bottleneck_inj}
\end{table}

\section{Design of a multistage collimation system}
\label{stages}
\subsection{Putting together the requirements to define a design work flow}

For the collimation system to fulfill the required cleaning goals, one must ensure that:
\begin{itemize}
\item[(1)] the aperture bottlenecks of the accelerators are shielded by a collimation system that sits at lower normalized apertures such that, for all loss scen\-arios, primary beam losses hit first collimators;
\item[(2)] the total energy carried by the beam, \ie out-scattered beam particles
  and the secondary prod\-ucts of beam particles' interactions with the collimator
  matter, is efficiently absorbed by the collimation region, with tolerable leakage
  to sensitive equipment, notably the protected bottlenecks, for any relevant loss
  scenarios. The cold magnets need special assessment as they are subject to quenching;
\item[(3)] the collimators themselves and other equipment installed in the
  dedicated regions must with\-stand, without damage, beam losses for different
  design scenarios;
\item[ (4)] the collimation contribution to machine impedance must be tolerable to the extent that the high-intensity
  beams remain stable. Any other considerations related to the operation with collimators at small gaps, \ie below the machine aperture, must also be taken into account: beam losses, operational alignment tolerances, impact on beam cuts and luminosity, ... .
\end{itemize}
Consistently to the notation introduced above, the item (2) can be expressed also by saying that we must operated above the quench limit of superconducting magnets for the design loss scenario when the beam experience the minimum allowed lifetime at full intensity. 


The last aspect in the list above is particularly critical when it comes to the design of collimators and absorbers. For more details, see~\cite{ab}, where considerations on material choices are discussed. The complete design of a complex system such as that of the LHC requires several steps and iterations across different domains, extending well beyond the field of accelerator physics. These aspects are not addressed in a fully comprehensive way in this document, although the key requirements will be mentioned.

It is important to stress that designing a collimation system from scratch is an iterative process, requiring multiple revisions before converging to a satisfactory solution. The initial understanding of the machine aperture and the cleaning specifications can be used to determine a first collimation layout and sets of transverse settings. These are then used as inputs in detailed simulations to calculate the collimation cleaning performance and compare it to the specified targets. If the performance is adequate, the technical design of the individual collimators can begin. In case of issues, improvements must be envisaged, either by upgrading the system layouts or by modifying the optics or designs of other accelerator components, depending on the limitations encountered — for example, increasing the aperture at the bottlenecks if additional margins are needed.

\subsection{Beam cleaning specifications}

Let us summarize the numbers introduced for the~LHC to illustrate how, in the initial design phase, one can produce quantitative specifications for the target cleaning performance. At the minimum allowed beam lifetime of $0.2\Uh$, and at full intensity with 7~TeV beams, one obtains a proton loss rate of $4.4\times10^{11}~\text{protons}/\UsZ$, corresponding to about 500\UkW. Using the quench limit of superconducting magnets, approximated by the expressions in \Eref{q}, one can derive the following target for the local cleaning inefficiency at 7~TeV:
\begin{equation}
\tilde\eta_\text{c}\le\frac{1}{10000} {\rm [1/m]}~.
\label{cleanSpecs}
\end{equation}
This target must be respected for all cold magnets around the ring.


This target must be respected for all cold magnets around the ring. Verification of the collimation system’s cleaning efficiency requires precise simulation tools, which model halo particle dynamics, interactions with collimator material, the ring aperture, and other relevant factors. While modern tools can combine particle tracking with energy deposition calculations in superconducting coils, the simplified approach of optimizing $\tilde\eta_\text{c}$ in Eq.~(\ref{cleanSpecs}) remains valuable for early-stage design. Proton losses per unit length can be simulated with fast and accurate setups~\cite{grd}, providing an essential tool for design optimization, while final validation relies on more detailed energy deposition simulations~\cite{fc}.

\subsection{Single-stage collimation}

\begin{figure}
  \centering
  \includegraphics[width=110mm]{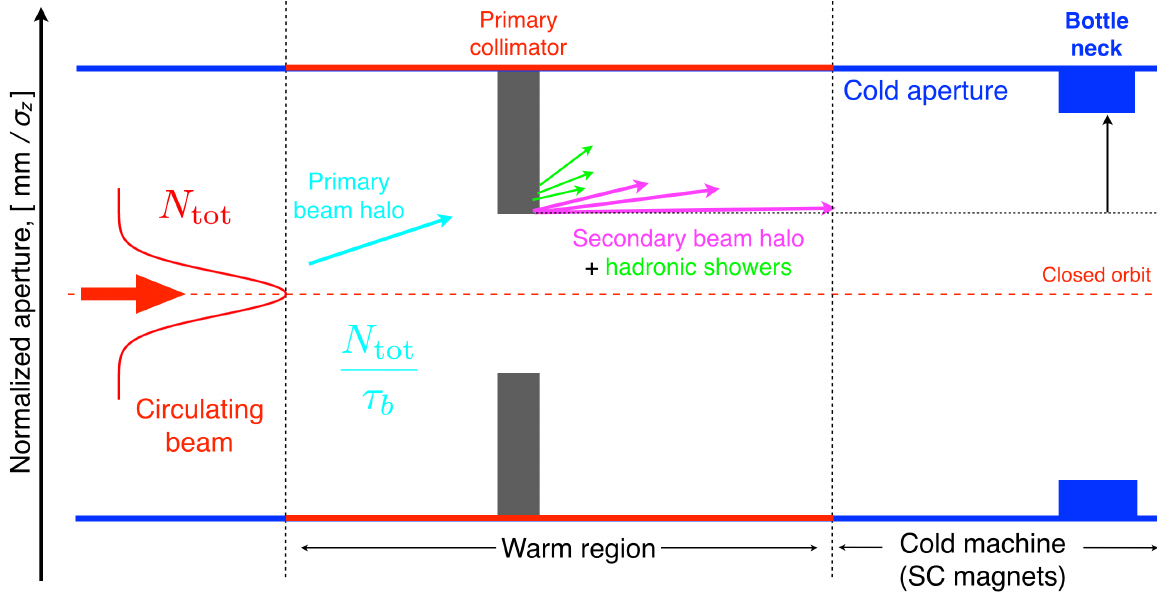}
  \vspace{-0.2cm}
  \caption{Single-stage collimation system: SC, superconducting}
  \label{fig_single}
\end{figure}


Designing a collimation system for a super-collider like the LHC requires finding an optics solution and a collimator arrangement that keeps losses in cold magnets below quench limits for all design loss rates. Consider a machine with a single cold bottleneck in the $z$ plane, as in \Fref{fig_single}. In an ideal case with no beam losses, collimation would not be needed if the minimum machine aperture were safely distant from the circulating beam core (red in \Fref{fig_single}). In practice, halo particles drift outward due to various loss mechanisms and eventually hit the aperture if not intercepted. The deposited energy at the~bottleneck then depends on the primary beam loss rate, $N_{\text{tot}}/\tau_\text{b}$.

A simple single-stage collimation system can be built by placing a primary collimator (TCP; ``target collimator, primary'') to intercept primary beam losses. Preferably, collimators are located in warm regions, away from superconducting magnets. The collimator jaws must be set at a normalized aperture smaller than that of the machine bottleneck, $n_{\rm TCP} < n_{\rm BTNK}$. In an ideal case, if the TCP were a perfect absorber, a single collimator could stop all primary particles on first passage. Because halo particle amplitudes mix due to betatron motion, a single jaw can protect the aperture against slow, diffusive losses. However, submicrometre impact parameters at first passage~\cite{seidel}, combined with jaw flatness and surface roughness, increase inefficiency, as multiple turns are required before particles accumulate enough interactions with the TCP.

\begin{figure}
  \centering
  \includegraphics[width=130mm]{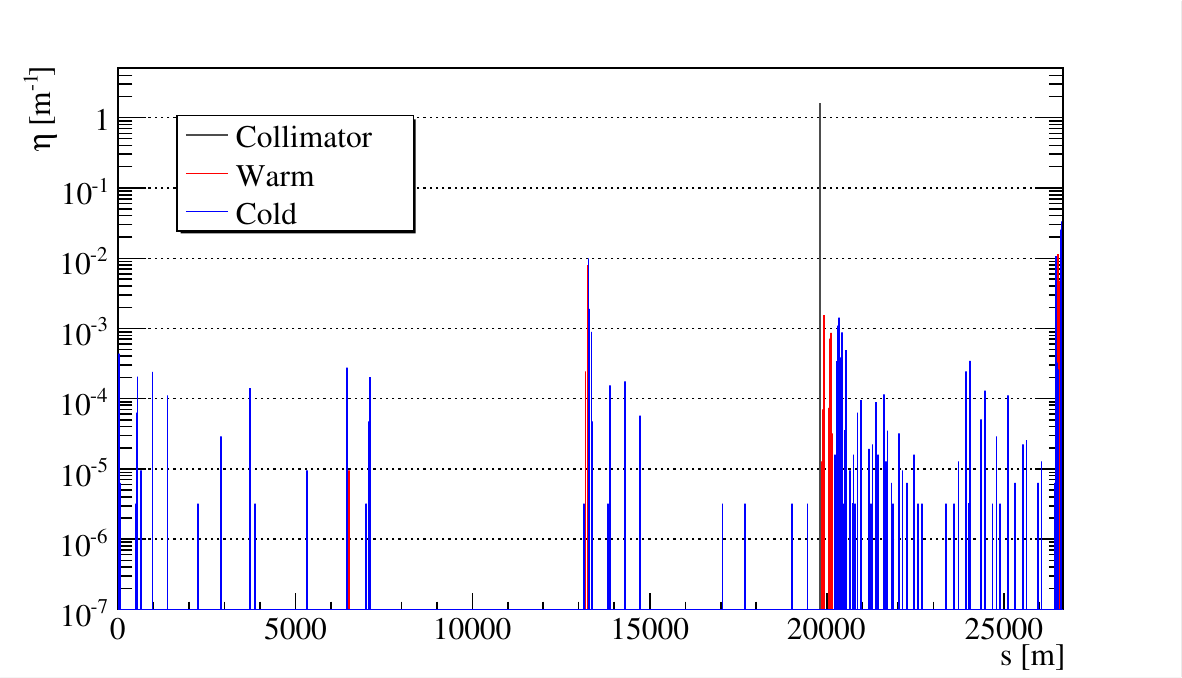}
  \vspace{-0.2cm}
  \caption{Simulated cleaning inefficiency at the LHC for a single-stage
    collimation system achieved with one hori\-zontal primary collimator (TCP)
    located at the beginning of the LHC warm betatron cleaning insert. The
    position of the existing primary collimators, \ie $s = 19.8\Ukm$, is used. {\it Courtesy of D. Mirarchi}.}
  \label{fig_ss-cleaning}
\end{figure}

\begin{figure}
  \centering
  \includegraphics[width=110mm]{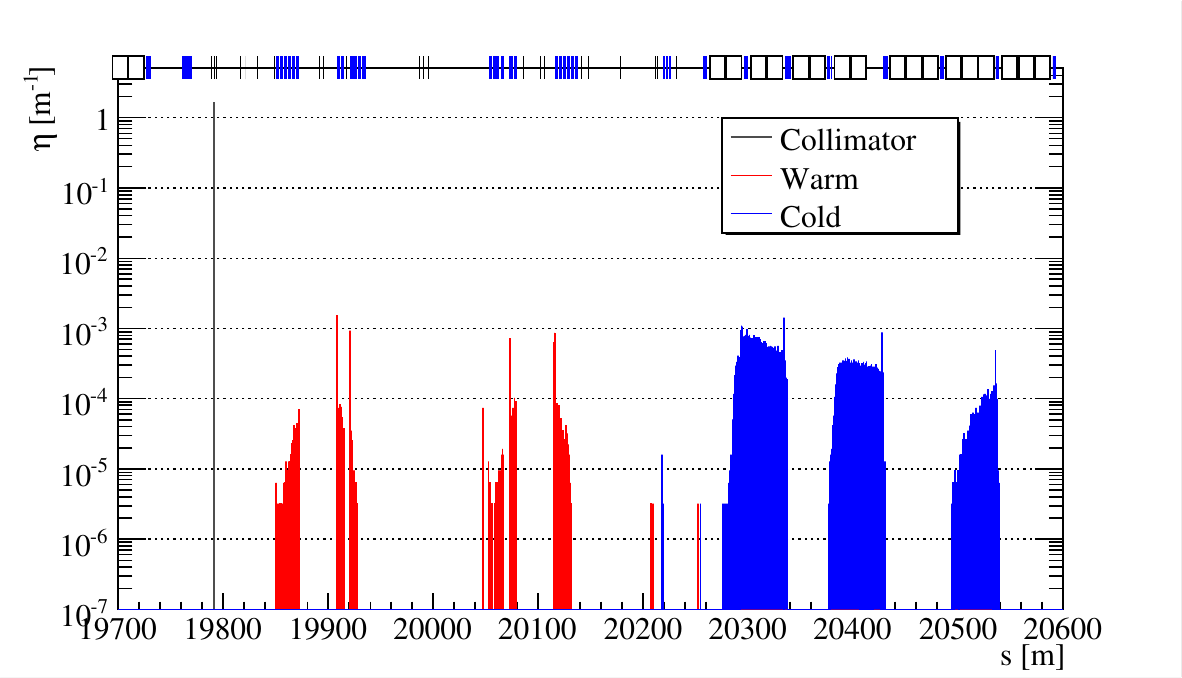}
  \includegraphics[width=110mm]{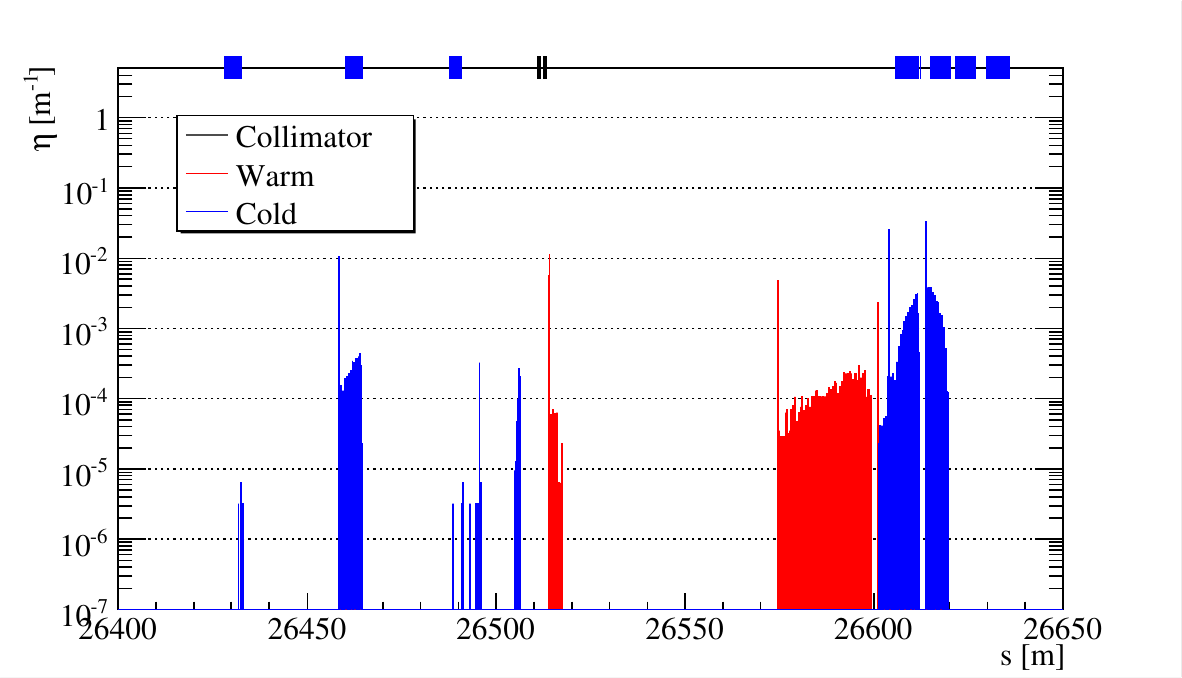}
  \vspace{-0.2cm}
  \caption{Enlargement of \Fref{fig_ss-cleaning}
    in the regions immediately downstream of the cleaning insertion (top) and
    upstream of the ATLAS experiment (bottom). {\it Courtesy of D. Mirarchi}.}
  \label{fig_ss-cleaning_z}
\end{figure}

For realistic machine parameters, the single-stage system of \Fref{fig_single} does not provide sufficient halo cleaning. Halo protons scattered by the jaw leave at larger normalized amplitudes and modified energies, forming a secondary beam halo that may be lost elsewhere in the machine. Hadronic and electromagnetic shower products are also not contained, potentially reaching sensitive elements without additional downstream collimators or absorbers.
The cleaning performance of this system was simulated assuming a horizontal TCP in the current LHC betatron cleaning insertion. The simulation tools of Ref.~\cite{grd} track protons through the magnetic lattice and their scattering in collimator materials. Figs.~\ref{fig_ss-cleaning} and \ref{fig_ss-cleaning_z} show the predicted local cleaning inefficiency, $\eta$, as a function of the longitudinal coordinate $s$, using the complete LHC layout and aperture model. All collimators other than the single primary TCP are ignored.

In the plots, black peaks indicate TCP losses, blue peaks indicate cold-magnet losses, and red peaks indicate warm-element losses. Zooming in near interaction points (\Fref{fig_ss-cleaning_z}), cold losses reach inefficiency levels of up to $0.01,/\UmZ$, at least two orders of magnitude above the target in Eq.~(\ref{cleanSpecs}). Multiple loss spikes also occur around the ring (\Fref{fig_ss-cleaning}). This demonstrates that a single-stage collimation system is inadequate for high-intensity superconducting machines such as the LHC.

\subsection{Multistage collimation}
\label{sec:multistage}

\begin{figure}
  \centering
  \includegraphics[width=110mm]{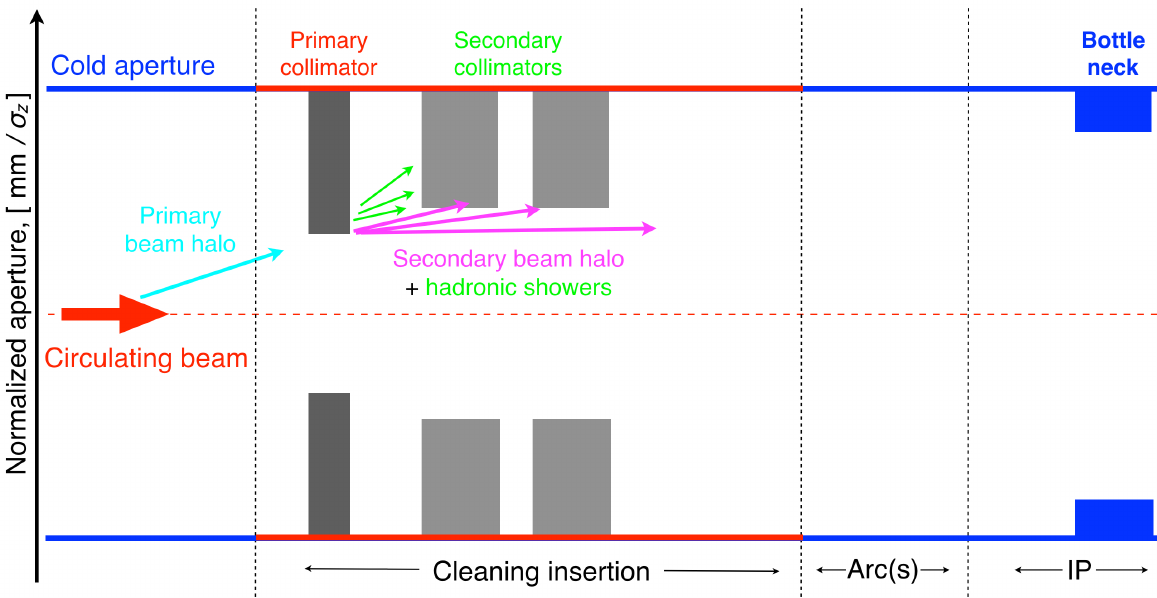}
  \vspace{-0.2cm}
  \caption{Two-stage beam collimation system, obtained by adding a set
    of secondary (TCS) collimators to the single-stage cleaning system of
    \Fref{fig_single}. IP, interaction point.}
  \label{fig_2stages}
\end{figure}

The performance of a single-stage system can be improved by adding downstream collimators to intercept secondary halo particles and the products of primary proton interactions with the TCPs, as shown in \Fref{fig_2stages}. These secondary collimators (TCSs) are typically longer than TCPs to maximize absorption of out-scattered particles. Their normalized apertures must be larger than the TCPs’ to maintain the \emph{collimation hierarchy}; if a TCS became the primary restriction, the system would perform like the inadequate single-stage setup discussed earlier. At the same time, TCS apertures should be small enough to efficiently capture particles scattered from the TCPs. Optimally, the hierarchy condition $n_{\rm TCP} < n_{\rm TCS} < n_{\rm BTNK}$ is respected, though this is not strictly required in all cases.

\begin{figure}
  \centering
  \includegraphics[width=70mm]{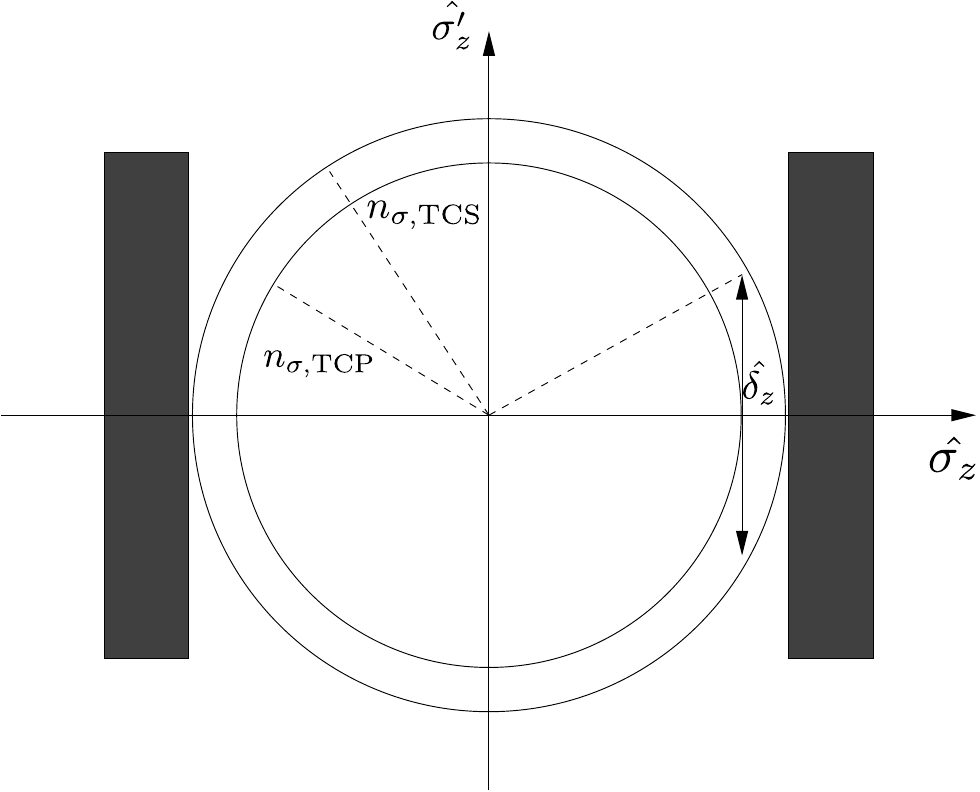}
  \vspace{-0.2cm}
  \caption{Normalized phase space with the circumferences radii
    $n_{\sigma, {\rm TCP}}$ and $n_{\sigma, {\rm TCS}}$. A normalized kick
    $\hat{\delta'}$, as in \Eref{d}, is necessary for halo particles impinging on the TCP to reach the TCS aperture.}
  \label{fig_delta}
\end{figure}

\begin{figure}
  \centering
  \includegraphics[width=80mm]{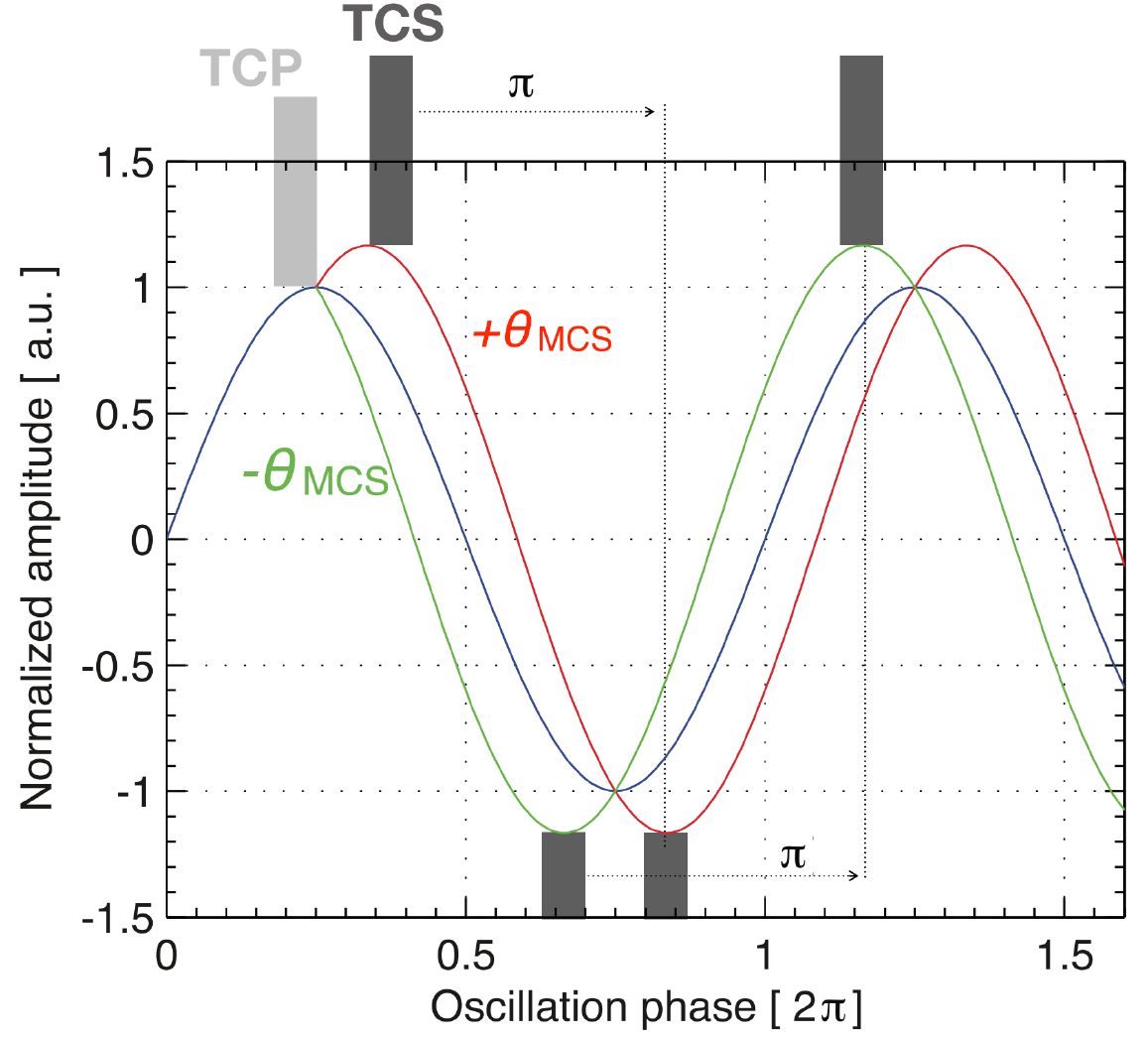}
  \vspace{-0.2cm}
  \caption{Qualitative definition of optimum locations for secondary
    collimators in a two-stage system, in which TCSs must intercept beam particles
    out-scattered at the primary collimators. In this one-dimensional model, two
    phase locations exist, where the amplitudes caused by multiple Coulomb
    scattering are a maximum for the two signs of the scattering angle,
    $\pm\theta_{\rm MCS}$. Two more TCSs at $\pi$ are added for redundancy.}
  \label{fig_phases}
\end{figure}

Figure~\ref{fig_delta} shown, in a normalized phase-space plot, the aperture in unit $\sigma_z$ of primary and secondary collimators. Collimator jaws are shown by dark-gray boxes at the zero-phase corresponding to the primary collimators. From \Fref{fig_delta}, one can calculate the kick of particles impinging on the TCP necessary to reach the amplitude of the TCS as
\begin{equation}
  \hat{\delta'}=\frac{\delta'}{\sigma'}
  =\sqrt{n_{\sigma, {\text{TCS}}}^2-n_{\sigma, {\text{TCP}}}^2}~,
  \label{d}
\end{equation}
where $\sigma'=\sqrt{\epsilon/\beta}$ is the r.m.s.~divergence. Such a kick is
typically accumulated after multiple passages through the TCP.
For a given TCS--TCP retraction, the longitudinal positions of the
TCS collimators must be optimized to intercept secondary halo particles. This is illustrated in \Fref{fig_phases} for a one-dimensional case. This condition is respected at betatron phase advances, where the multiple Coulomb  scattering angle translates into maximum offsets in the collimation plane.

The problem of optimum phase locations for a two-stage collimation system
is worked out in detail in Ref.~\cite{jbj} for the generic case of 2D betatron halos. Finding a solution is more complicated
than appears in \Fref{fig_phases} because scattering occurs in all
directions. A one-dimensional model is thus not adequate. However, it
can be demonstrated that an arrangement of primary and secondary collimators
in three planes (horizontal, vertical and skew) can be found to ensure
satisfactory multiturn cleaning \cite{jbj}.



Detailed performance studies of the two-stage cleaning system for the~LHC were conducted during the design phase~\cite{cham2005}. While this scheme efficiently shields the aperture from transverse halo losses, it cannot fully absorb hadronic shower products before they reach cold magnets downstream. Moreover, a two-stage system confined to a single insertion does not provide adequate local protection for critical bottlenecks, such as the triplet magnets around experiments, which are most critical during the squeeze. Consequently, the LHC collimation system evolved into a \emph{multistage system}, adding tertiary collimators (TCTs) in front of critical bottlenecks and shower absorbers (TCLAs) in the warm cleaning inserts, in addition to TCPs and TCSs.

\begin{figure}
  \centering
  \includegraphics[width=110mm]{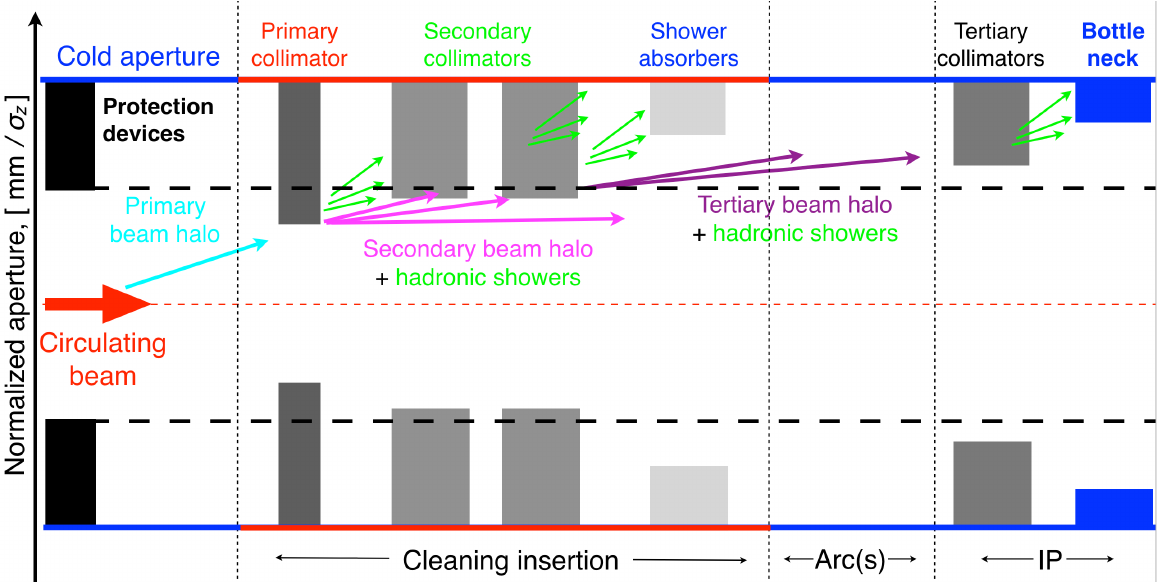}
  \vspace{-0.2cm}
  \caption{Key elements of the LHC multistage collimation system: IP, interaction point}
  \label{fig_multi}
\end{figure}


In addition, dump protection devices are added to the hierarchy to shield the machine in case of fast kicker failures, such as the asynchronous dump scenario introduced above. Similarly, injection protection devices are required during the injection process to protect the machine in case of injection kicker failures. This produces the final multistage collimation system shown in \Fref{fig_multi}. Although injection and dump protection devices are designed using a different technique, focusing on beam absorption during single-pass failures, their hierarchy is set consistently with the ring collimation system, since the protected aperture and device settings are defined with respect to the same circulating-beam orbit. For the~LHC, denoting the protection device aperture as ``TCDQ,'' the following multistage hierarchy is defined:
\begin{equation}
    n_{\rm TCP}<n_{\rm TCS}<n_{\rm TCDQ}<n_{\rm TCT}<n_{\rm BTNK}<n_{\rm TCLA}~.
    \label{EqHierarchyFull}
\end{equation}
This condition is of course specific to the LHC case. Notably, TCT and TCLA collimators are based on metallic jaws for maximum absorption, as discussed in the next section. Thus, they also need to be protected in case of fast failures contrary to TCP and TCS collimators that are more robust. 

\begin{figure}
  \centering
  \includegraphics[width=120mm]{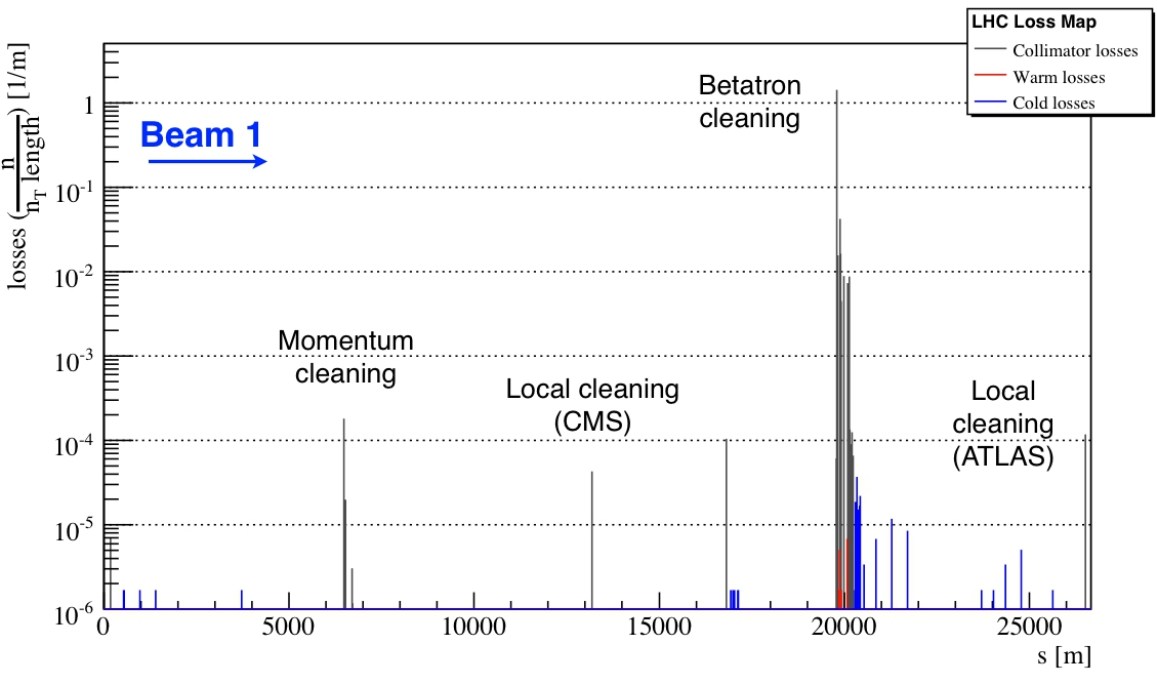}
  \vspace{-0.2cm}
  \caption{Local cleaning inefficiency as a function of $s$ for the final
    collimation system of the LHC run~I. Loss distributions are simulated for
    the LHC beam~1 at $7\UTeV$ for a perfect machine, with the collision
    optics squeeze to $\beta^*=0.55\Um$ in IR1 and IR5. {\it Courtesy of
      D.~Mirarchi}.}
  \label{fig_cl}
\end{figure}

\begin{figure}
  \centering
  \includegraphics[width=120mm]{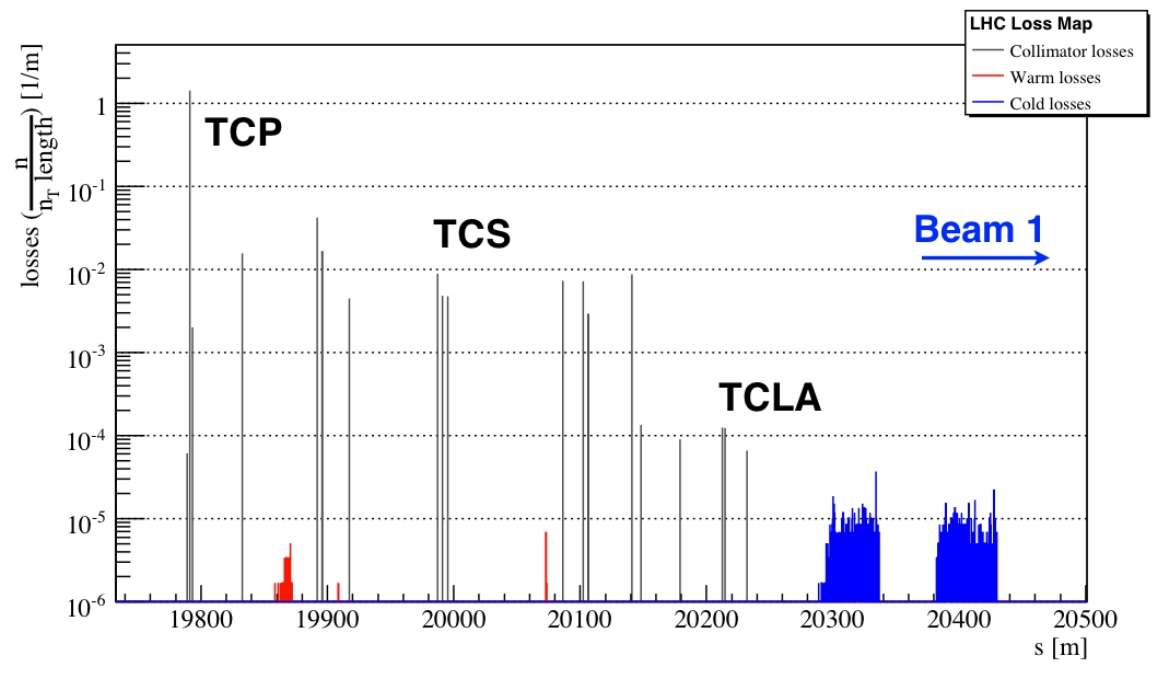}
  \vspace{-0.2cm}
  \caption{Enlargement of the IR7 region of the cleaning inefficiency plot of
    ~\Fref{fig_cl}. Labels indicate the approximate locations of the three
    families of collimator in IR7. TCLA, target collimator long absorber; TCP, target collimator (primary); TCS, target collimator (secondary).}
  \label{fig_cl_z}
\end{figure}

The cleaning performance of the final LHC collimation system
\cite{finalColl} is shown in ~\Fref{fig_cl}. While the~system is described in detail in the next section, the simulations are shown here for
a direct comparison with the single-stage system. The insertion regions
(IRs) where the largest losses occur are the~betatron (IR7) and momentum (IR3) cleaning, ATLAS (IR1) and CMS (IR5). This simulation is for beam 1 (B1), nominally
$7\UTeV$,  in collision conditions. An enlargement of the
loss map around the~betatron cleaning insert is shown in
\Fref{fig_cl_z}. For a perfect machine, cold losses are now below $\sim10^{-5}$. The~highest peaks are localized in the dispersion suppressor regions
downstream of IR7.

\section{The final LHC collimation system design}
\label{lhc-coll}
It has been shown that the LHC features a complex and distributed system is needed to achieve the~excellent halo
cleaning and robust operation below quench limits.
In this section, the
collimation layout is presented and the collimator design is reviewed. Operational challenges for the collimation at the LHC are then introduced, presenting
the solutions produced to set the system up for optimum performance in all
operational phases.
Some aspects of the achieved operational performance are also discussed.

\subsection{LHC ring collimation layout}
\begin{figure}
  \centering
  \includegraphics[width=120mm]{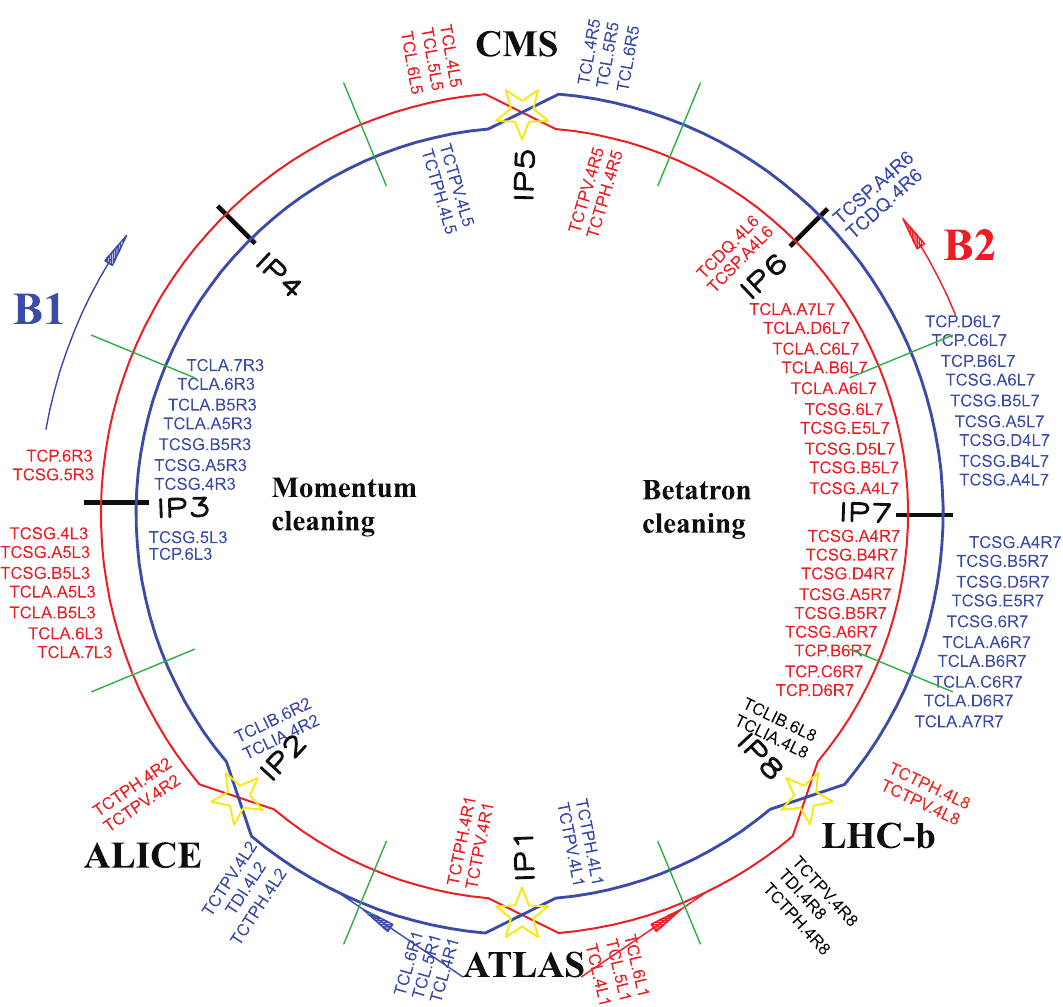}
  \vspace{-0.2cm}
  \caption{Layout of the LHC, showing the collimator locations around the
    ring}
  \label{figLayout}
  \vspace{-.3cm}
\end{figure}

\Figure[b]~\ref{figLayout} shows the LHC layout and the positions of the collimators around the
ring. A list of collimator types, with a
description of their functionality (primary, secondary, \etc) and key
collimator properties is given in Table~\ref{tabList}. Including the dump
protection block (target collimator dump quadrupole, TCDQ) and the injection protection collimator (target dump, injector TDI), the system deployed for the 2015 LHC run comprises 110 movable collimators installed in the LHC ring and its transfer lines. By the time of writing, the collimation layouts underwent several upgrades and have been significantly improved (see for example \cite{Arduini:2024kdp} for a complete review of the collimation upgrades for the Run~3 starting in 2022). The~key features of the collimation system, the described functionalities and the considerations made in this lecture remain however valid for the upgraded system.

\begin{table}
  \caption{List of movable LHC collimators for Run~2 (2015-2018). CFC, carbon fibre composite; H, horizontal; S, skew; V, vertical.}

  \begin{center}
    \begin{tabular}{lllrl}
      \hline\hline
      \textbf{Functional type }    & \textbf{Name}    & \textbf{Plane} & \textbf{Number} & \textbf{Material} \\
      \hline
      Primary IR3         & TCP     & H     & 2  & CFC \\
      Secondary IR3       & TCSG    & H     & 8  & CFC \\
      Absorbers IR3       & TCLA    & H,V   & 8  & W alloy\\
      Primary IR7         & TCP     & H,V,S & 6  & CFC \\
      Secondary IR7       & TCSG    & H,V,S & 22 & CFC \\
      Absorbers IR7       & TCLA    & H,V   & 10 & W alloy\\
      Tertiary IR1/2/5/8  & TCTP    & H,V   & 16 & W \\
      Physics debris absorber& TCL     & H     & 12  & Cu/W alloy \\
      Dump protection     & TCSP    & H     & 2  & CFC \\
                          & TCDQ    & H     & 2  & C \\
      Injection protection (lines)    & TCDI    & H,V   & 13 & CFC \\
      Injection protection (ring)   & TDI     & V     & 2  & C \\
                          & TCLI    & V     & 4  & CFC \\
                          & TCDD    & V     & 1  & CFC \\
      \hline\hline
  \end{tabular}

  \end{center}

 \label{tabList}
\end{table}


Halo collimation is achieved by the multistage cleaning system introduced in \Sref{stages}, which comprises three stages in IR3 (momentum cleaning) and IR7 (betatron cleaning). In each stage, primary collimators (TCPs) closest to the beam are followed by secondary collimators (TCSs) and active absorbers (TCLAs). For optimal performance, halo particles should first hit a TCP, while TCSs intercept only secondary halo particles scattered from upstream collimators. TCPs and TCSs, being closest to the~beam and subject to large losses, are made of carbon-fiber composite (CFC) for high robustness and to withstand potential beam impacts during failures. TCLAs catch tertiary halo particles scattered from the TCSs, as well as showers from upstream collimators. Made of a tungsten alloy, TCLAs absorb as much energy as possible but are less robust than CFC collimators and should never intercept primary beam losses. The collimator hierarchy is set to ensure this condition is respected under all operating conditions.



In addition to the dedicated collimation regions in IR7 and IR3, collimators are installed in most other IRs. A pair of tertiary collimators with embedded orbit-measurement pick-ups (TCTPs), made of tungsten alloy, are placed about $150\Um$ upstream of the collision points for all experiments, with one in the horizontal plane (TCTPH) and one in the vertical (TCTPV) planes. They provide local protection of the quadrupole triplets in the final focusing system, which are the limiting cold apertures during physics operation, and help reduce experimental background. Downstream of the high-luminosity experiments, ATLAS and CMS, three long collimators (TCLs) per beam intercept collision debris. At beam extraction in IR6, dump protection collimators protect against miskicked beams, while injection protection collimators are installed in IR2 and IR8.

During the 2013–2014 LHC long shutdown, 18 new collimators incorporating beam position monitor (BPM) pick-ups~\cite{carra} were installed. These replaced the TCSGs (secondary graphite collimators) in IR6 and the tertiary collimators in all experiments: these locations are critical for orbit control. News collimators enhanced significantly the LHC performance~\cite{roderikBeta}. This new design is retained for all future upgrade designs. 

\subsection{Optics and layout of cleaning inserts}
\begin{figure}
  \centering
  \includegraphics[width=119mm]{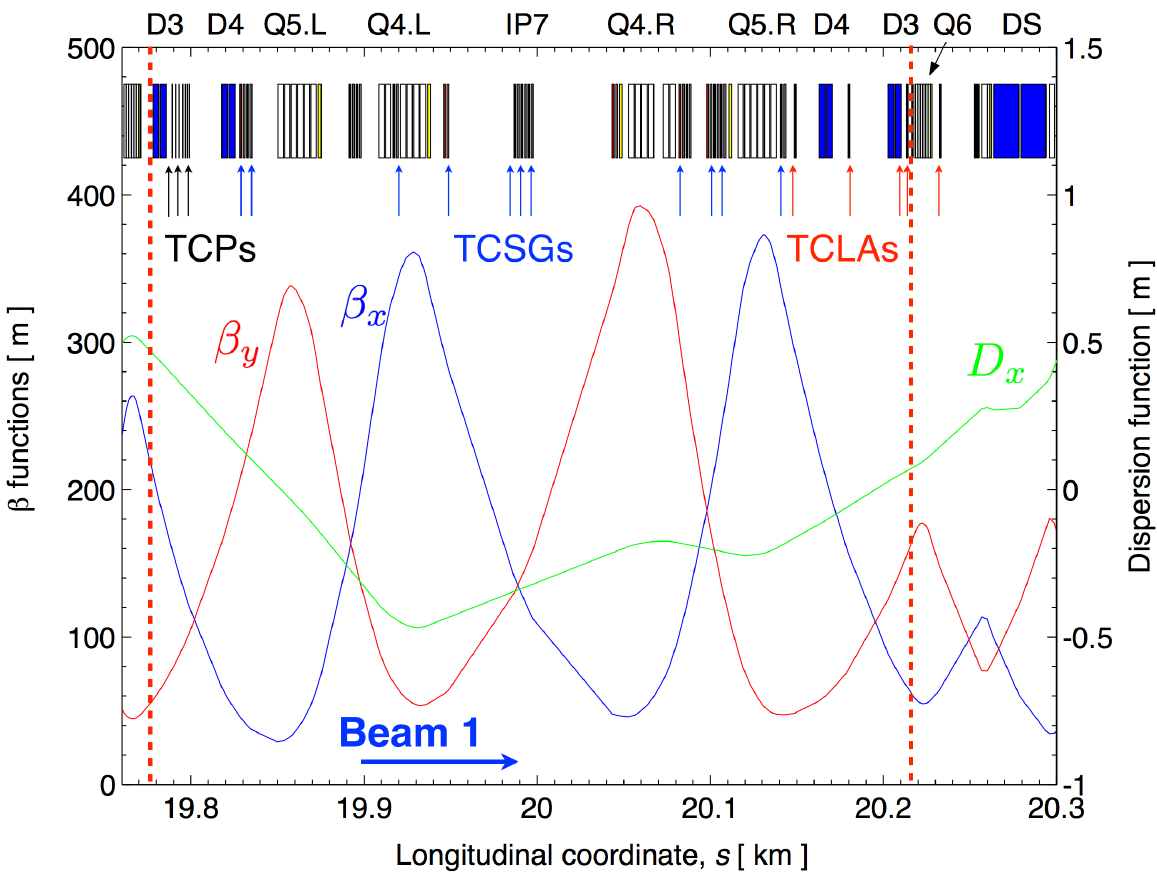}
  \vspace{-0.2cm}
  \caption{Betatron ($\beta_x$, $\beta_y$) and dispersion ($D_x$) functions
    as a function of $s$
    in the LHC betatron cleaning insertion IR7. The main layout elements
    are also shown: quadrupoles (white boxes), dipoles (blue), and collimators
    (black). Vertical arrows indicate the installed collimators: 3 TCPs, 11 TCSGs;
    5 TCLAs. Vertical red dashed lines indicate the limits of the warm regions (Q6 magnets at either side of IP7 are the first cold magnets). D, dipole magnet; IP, interaction point; L, left; R, right; Q, quadrupole magnet; TCLA, target collimator long absorber; TCP, target collimator (primary); TCSG, target collimator (secondary, graphite).}
  \label{figLayoutIR7}
  \vspace{-.3cm}
\end{figure}

\begin{figure}
  \centering
  \includegraphics[width=119mm]{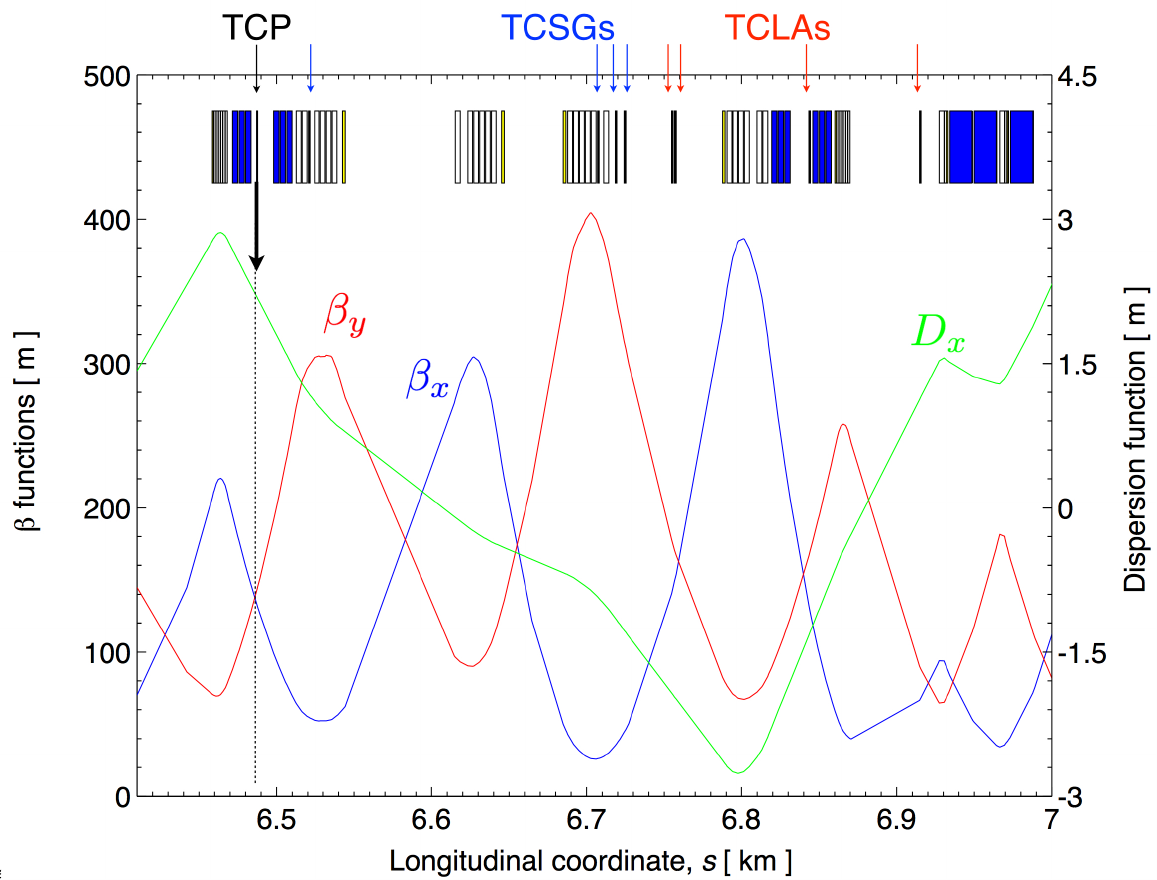}
  \vspace{-0.2cm}
  \caption{Betatron ($\beta_x$, $\beta_y$) and dispersion ($D_x$) functions
    as a function of $s$
    for B1 in the LHC momentum cleaning insertion IR3. Vertical arrows indicate
    the installed collimators: 1 TCP, 4 TCSGs;
    4 TCLAs. The main layout elements
    are also shown: quadrupoles (white boxes), dipoles (blue), and collimators
    (black). TCLA, target collimator long absorber; TCP, target collimator (primary); TCSG, target collimator (secondary, graphite).}
  \label{figLayoutIR3}
  \vspace{-.3cm}
\end{figure}

The optics and layout of the betatron and momentum cleaning inserts are
shown in Figs.~\ref{figLayoutIR7} and \ref{figLayoutIR3}, respectively~\cite{lhc}. In
both inserts, four \emph{dog-leg} dipoles, called D4 and D3, are placed symmetrically on either side of the `IP7', and are used to enlarge the
beam--beam separation from $194\Umm$ to $224\Umm$, making more transverse space
for collimators. The two D4 magnets also delimit the
${\approx}500\Um$
long warm insert, which  comprises the warm quadrupoles Q4 and Q5. The Q6
quadrupoles on either side of the D4 dipoles are the first superconducting
magnets before the beam enters the cold arc.

In IR7, three primary collimators intercept horizontal, vertical, and skew halos. They are located between the D3 and D4 dipoles, upstream of the warm insertion for each beam, which maximizes the~length of the warm section downstream of the primary loss location. This placement also prevents neutral particles produced at the TCPs from having a direct line of sight to downstream cold magnets, thanks to the tilted orbit between D3 and D4 where TCPs are installed. A similar scheme is used in IR3, where only a single horizontal TCP is required. It is placed at a location with large normalized dispersion, $D_x/\sqrt{\beta_x}$, to intercept particles with energy deviations. The IR3 primary collimator is set at larger transverse betatron amplitudes than in IR7, decoupling the functionalities of the two inserts. Typical transverse amplitudes, expressed in units of $\sigma_x$ as in \Eref{sz}, are 2.5–3 times larger than in IR7, ensuring that IR3 does not act as a betatron collimation system for particles with small energy errors.



The collimators in IR3 and IR7 are shown in Figs.~\ref{figLayoutIR7} and \ref{figLayoutIR3} as black boxes. IR7 uses eleven TCS collimators, while IR3 uses four, since collimation occurs in a single plane. Five TCLA absorbers are installed in IR7 and four in IR3. All devices—TCPs, TCSGs, and TCLAs (see Table~\ref{tabList})—are two-sided. While a one-sided collimator might suffice for a multiturn cleaning process, two-sided devices are essential for precise beam alignment.

\begin{figure}
  \centering
  \includegraphics[width=\linewidth]{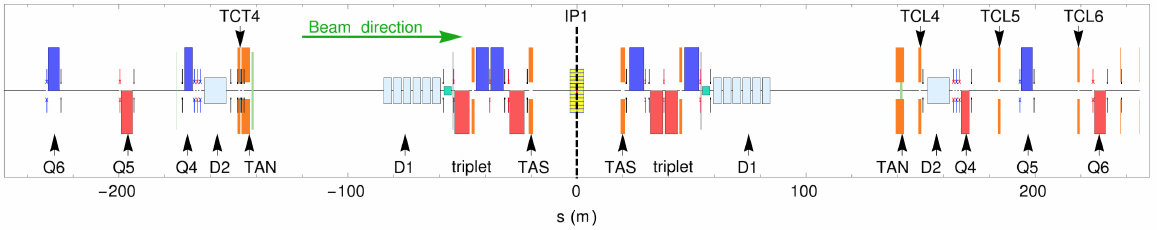}
  \vspace{-0.8cm}
  \caption{Layout elements around IR1 (ATLAS) in the 2015 configuration of the
    LHC collimation system. D, dipole; IP, insertion point; Q, quadrupole; TAN, target absorber (neutral); TAS, target absorber (secondary); TCL, target collimator (long); TCT, target collimator (tertiary). {\it Courtesy of R.~Bruce}.}
  \label{figIR}
  \vspace{-.3cm}
\end{figure}

The layout of IR1 (ATLAS) is shown in \Fref{figIR}, where a pair of horizontal and vertical TCTPs protects the triplet from incoming beam losses. Three TCL-type physics debris absorbers shield the~downstream magnets from collision products. IR5 (CMS) has an equivalent layout. In IR2 (ALICE) and IR8 (LHCb), TCL collimators are not needed, as the lower luminosity does not put the matching sections at risk of quenching.

In addition to movable collimators, 10 passive absorbers (TCAPs) are mounted in front of the~most exposed warm magnets of each collimation insertions: the D3 magnets downstream of the TCPs and the~first modules of the Q5 and Q4
quadrupoles. These fixed-aperture collimators dramatically reduce the~radiation doses to magnet coils, increasing their lifetimes by a factor of 10  or more (chapter 18 of \cite{lhc}).
\newpage
\subsection{Operational challenges and beam-based set-up}
\subsubsection{LHC operational cycle and recap. of machine configurations}
The main phases of the LHC operational cycle, which is
periodically run to prepare for periods of physics data acquisition
({\it stable beam} mode), are injection, energy ramp, betatron squeeze, and
preparation of collisions ({\it adjust}). The {\it squeeze}, in which the optics
around the interaction points are changed to reduce the colliding beam sizes. It can be executed at constant flat-top energy or it can (partially) take place during the energy ramp. In the squeeze, the
betatron function is enlarged at the inner triplets as required to reduce
the $\beta^*$ values, \ie the beta functions at the collision points.

The
LHC design value of $\beta^*$ for the high-luminosity points IP1 (ATLAS) and IP5
(CMS) is $55\Ucm$ for a beam energy of $7\UTeV$, limited by the available triplet
aperture. During LHC run~I, a $\beta^*$ value of $60\Ucm$ was achieved at $4\UTeV$.
The first year, 2015, of LHC Run~2 started with a $\beta^*$ value of $80\Ucm$
at $6.5\UTeV$, to ease recommissioning after the 2 year shutdown~\cite{beatCham2014},
but it finally achieved the smallest $\beta^*$ values of $25$~cm. If the squeeze is carried out during the collision process as a mean to control the~luminosity, it is referred to as {\it $\beta^*$ levelling}~\cite{Calia:2024paf}.
These excellent results were achieved thanks to a~better aperture than had been anticipated during the LHC design phases, which was also better than the~one
used to specify various LHC systems. For the scope of this lecture, it
is still useful to review the~system design by starting from the design values.

\subsubsection{Collimation settings strategy in the LHC operational cycle}

The LHC aperture was reviewed in \Sref{notation}; see Table~\ref{tab_bottleneck_inj}. The injection stored beam energy is already $22\UMJ$, \ie not only above the quench limit but also significantly above the damage limit of
metals \cite{ab}. Beam colli\-mation is therefore required in every phase of the LHC operational cycle, from
injection to collision. Particularly challenging are the dynamic phases (energy ramp, betatron squeeze, and change of orbit configurations), when collimator movements must be synchronized with other accelerator systems, such as power converters and radio-frequency. As discussed in Section~\ref{Sec:apert}, preserving optimum collimation settings required movable collimators to follow the beam envelop, \eg for primary collimators at normalized settings of 5--6~$\sigma_z$.

\begin{figure}
  \centering
  \includegraphics[width=115mm]{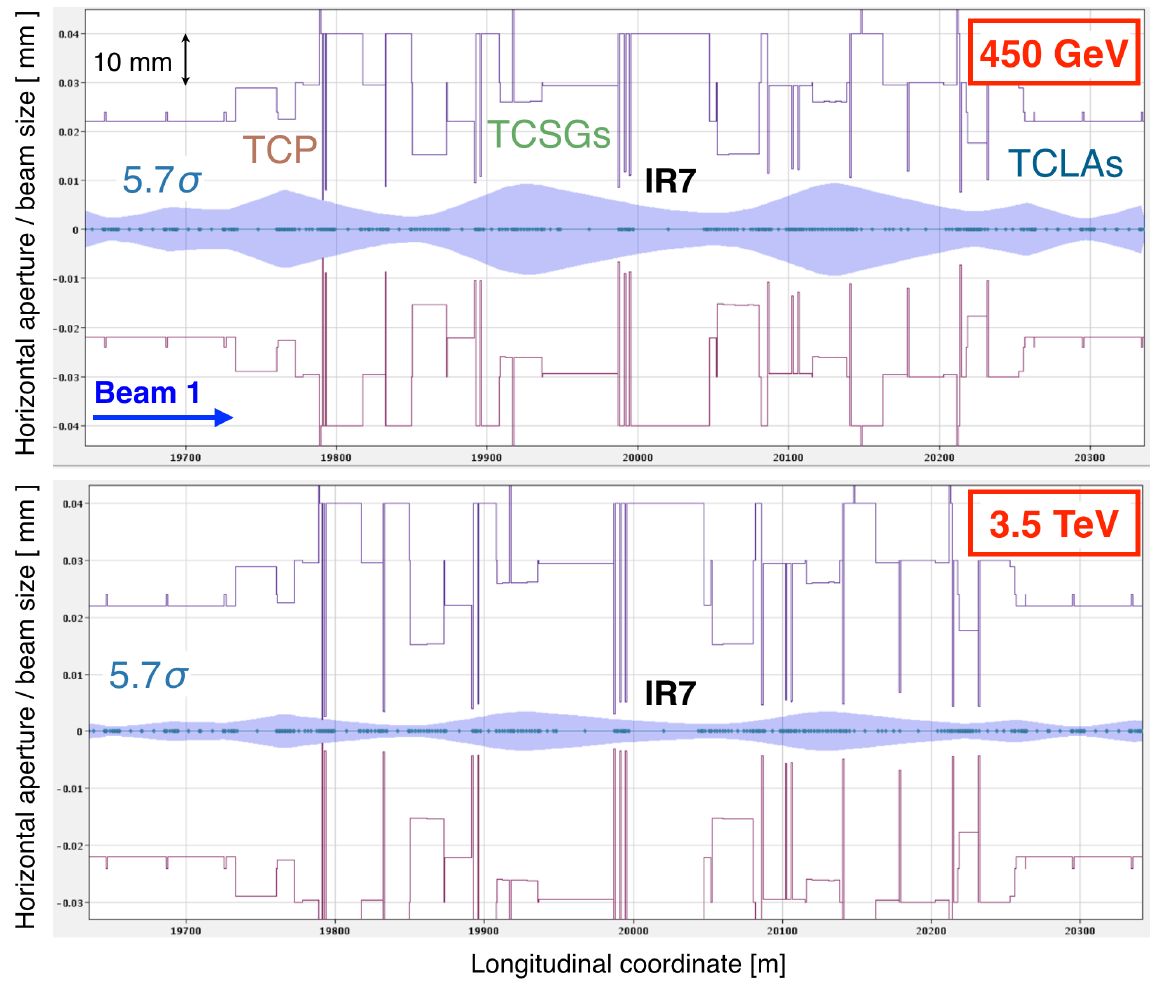}
  \vspace{-0.2cm}
  \caption{Horizontal aperture, collimator jaw positions (vertical bars) and
    $5.7\sigma$ beam envelope at (top) injection and (bottom) $3.5\UTeV$ in
    betatron cleaning (IR7) from the LHC on-line model application
    \cite{hb2010}. IR, insertion region; TCG, target collimator (graphite); TCLA, target collimator (long absorber); TCP, target collimator (primary).}
  \label{figIP7gaps}
\end{figure}
\begin{figure}
  \centering
  \includegraphics[width=105mm]{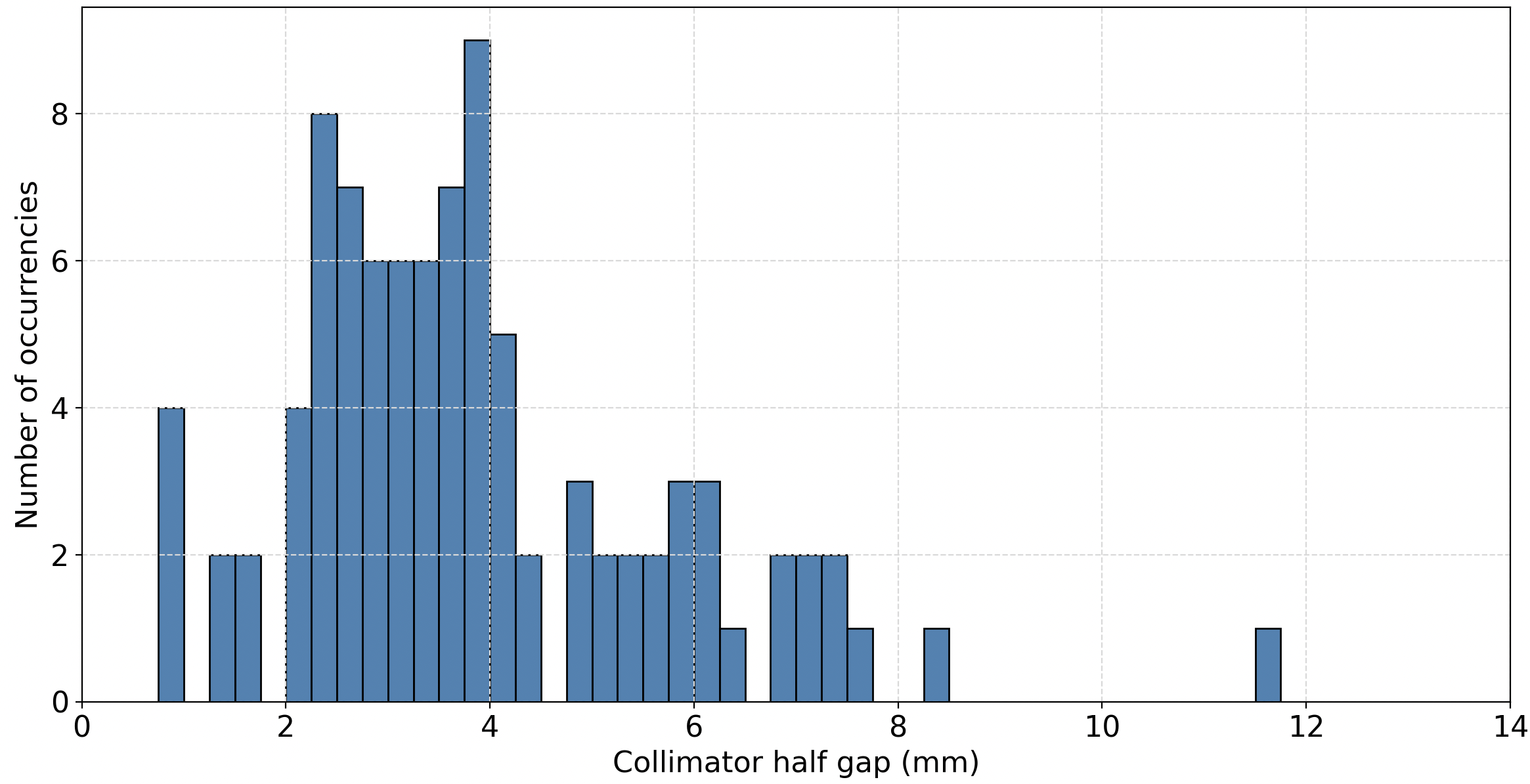}
  \vspace{-0.2cm}
  \caption{Distribution of collimator half gaps in operation at the LHC during 2025 at 6.8~TeV.}
  \label{fig_smallgaps}
\end{figure}

\Figure[b]~\ref{figIP7gaps} shows an example of collimator settings at injection (top) and $3.5\UTeV$ (bottom), taken from the operation configuration of the LHC 2010 run \cite{hb2010}. The horizontal beam envelope at
$5.7\sigma_x$, as defined by the TCP gaps, is shown, together with the values of the collimator half gap projected on the~horizontal plane at each collimator (magenta bars). The TCPs were kept at a normalized aperture of $5.7\sigma_z$ at all energies. The TCSGs were moved from $6.7\sigma_z$ to $8.5\sigma_z$, and the TCLAs were moved from $10.0\sigma_z$ to $17.7\sigma_z$. These relaxed top-energy settings were conceived to reduce the operational tolerances in the first year of the run \cite{hb2010} and were then subsequently tightened to improve the cleaning performance \cite{beta2012}, reaching $4.3\sigma_z$ in 2012. The present operation at 6.8~TeV features much tighter settings of $5.0\sigma_z$, thanks to a continuous optimization of the system~\cite{VanderVeken:2023ryw, Redaelli:2024glc}. The collimator gap values in millimetres, as used for the $6.8\UTeV$ operation in 2025 are shown in \Fref{fig_smallgaps}, where the transverse
clearance left by the IR7 primary collimators and the distribution of gaps are shown. The smallest half gap is below $1\Umm$.

\subsubsection{Beam-based set-up of LHC collimators}
The results Figs.~\ref{figIP7gaps} and \ref{fig_smallgaps} demonstrate further that movable-jaw collimators must be envisaged: the~gaps required at top energy to ensure optimum performance are not compatible with the
larger beam sizes at injection. Given the small gaps at top energy, collimators cannot be set deterministically but need to be aligned to the circulating beams, through custom-developed procedures referred to as {\it beam-based alignment} procedures. With beam sizes as small as $200\Uum$ and orbit offsets of up to 2--3\Umm at collimators, and in the presence of collimator survey errors of up to a few hundred micrometres, the~determination of optimum jaw positions can only be achieved through a series of measurements aimed at measuring the~required parameters.

\begin{figure}
  \centering
  \includegraphics[width=120mm]{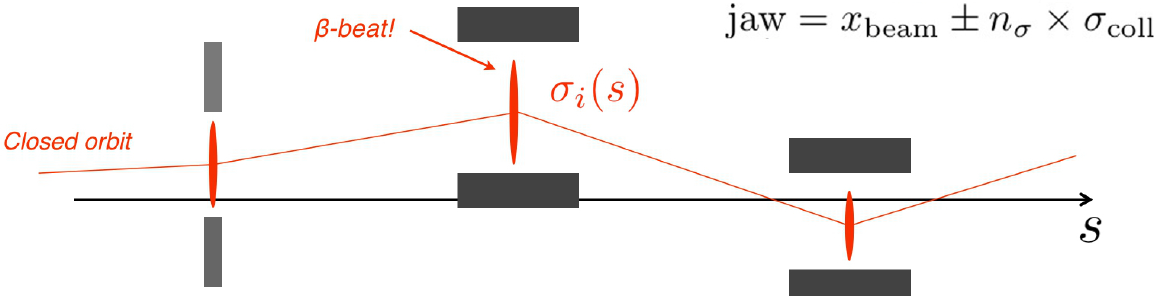}
  \vspace{-0.2cm}
  \caption{Scheme of orbit and beam sizes at various collimator locations. Measure {\it beam-based} values are used to ensure that the collimator hierarchy is respected.}
  \label{figAlign}
\end{figure}

\begin{figure}
  \centering
  \includegraphics[width=130mm]{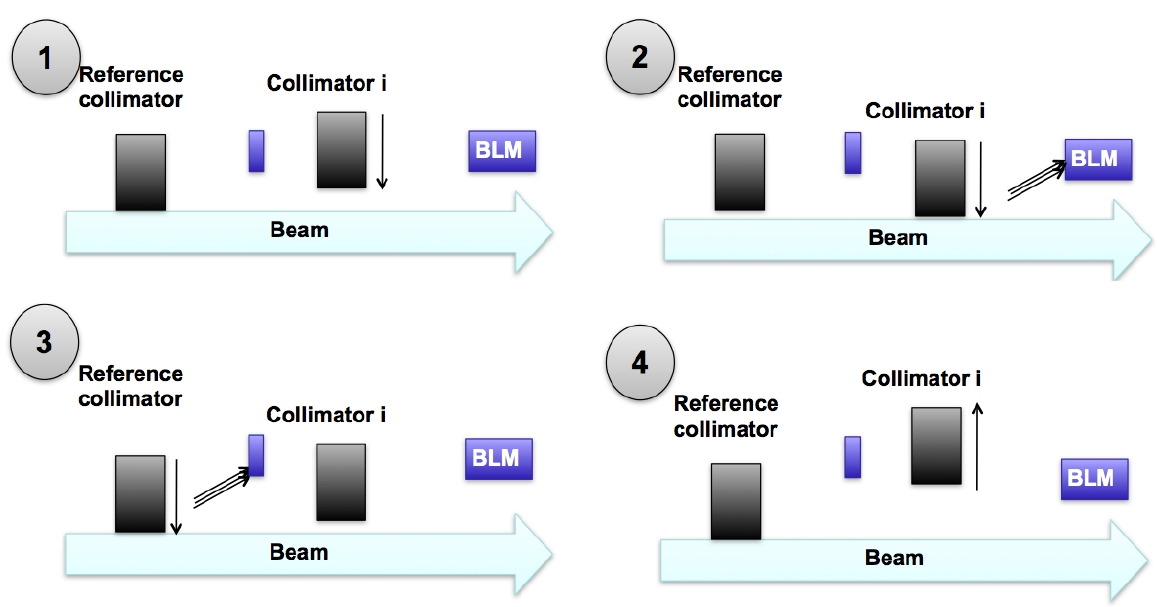}
  \vspace{-0.2cm}
  \caption{The collimator set-up procedure used to determine
    beam orbit and relative beam size to that of a reference collimator,
    for operational settings generation; BLM, beam loss monitor.}
  \label{figSetupProc}
  \vspace{-0.2cm}
\end{figure}

Deploying operationally the well-defined collimation setting hierarchy for the different families, see Eq.~(\ref{EqHierarchyFull}), involves knowing the beam orbit and beam size at each collimator, as shown in \Fref{figAlign}. Instead of nominal values, for the collimator setting generation beam-based values are used. These are established experimentally through a procedure (\Fref{figSetupProc}) that was established \cite{hb2010} based on ex\-peri\-ence gained with an LHC collimator prototype installed in the Super Proton Synchrotron (SPS) \cite{spsExp}.  

The beam halo is shaped with a reference collimator (1), typically a primary collimator, which is closed to a known half gap of $3-5\sigma$. This reference halo is used to cross-align other colli\-mators, by moving their jaws towards the beam in small steps of 5--20$\Uum$ until the halo is \emph{touched}, with symmetrical beam loss responses from either jaw (2). This gives the local orbit position. The reference collimator is then closed further (3) until it touches the halo again: this enables the gaps of the two collimators to be cross-calibrated. The average of the initial and final gaps of the reference collimator in units of $n_\sigma$ gives the normalized gap of the other collimator. Finally, the latter collimator is opened to its nominal settings (4). This ensures that the relative retraction with respect to the reference collimator is respected, even in the presence of different beta-beating at the two locations. 

This set-up procedure is precise but time consuming. During the initial commissioning in 2010, it was carried out manually for each collimator. An automated feedback system between collimator movements and the beam loss monitor
signal has been developed, enabling the set-up time to be improved significantly and dramatically reducing the number of spurious beam aborts from human errors. The~beam-based alignment procedure is supported by machine learning (ML) algorithms, whose presentation is beyond the scope of this lecturer. A detailed presentation can be found in Ref. \cite{thesisGV}, with the~latest ML developments and results available in~\cite{Azzopardi:2019bea, Arpaia:2020mvq}.

\subsubsection{Collimator setting generation for operation}
\label{sec:coll-sett}
Beam-based alignment must be done for each collimator in the ring, for every
relevant machine con\-figur\-ation (injection, top energy before and after squeeze,
collision). To minimize the risk of damaging the~collimators while approaching
them to the beams and to control beam losses in the process, the~alignment is carried out with the minimum intensity that
allows reliable orbit measurements, \ie with a few bunches of nominal
intensity. Let us now assume that the local orbit, $x_{\text{beam}}$, and beam size,
$\sigma_{\text{coll}}$, are calculated at every collimator in each discrete point
of the operational cycle.

\begin{figure}
  \centering
  \includegraphics[width=75mm]{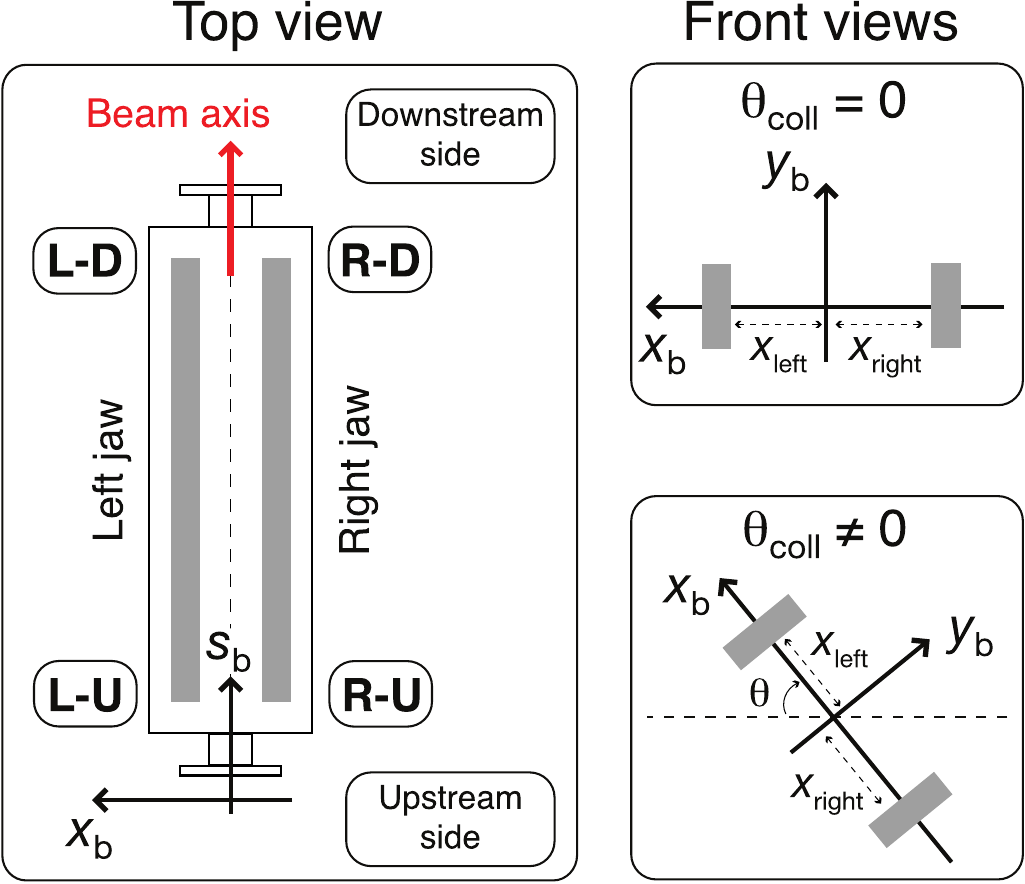}
  \vspace{-0.2cm}
  \caption{Top and front views of a collimator, with labels
    and naming conventions. Each jaw has two motors that move the jaws in the
    collimation plane: horizontal ($\theta_\text{coll}=0$), vertical
    ($\theta_\text{coll}=\pi/2$) or
    skew planes. D,~downstream; L, left; R, right; U, upstream.}
  \label{figNaming}
\end{figure}

While collimators are installed in a variety of azimuthal orientations (see
\Fref{figNaming}), the jaw move\-ment is in one dimension, along the collimator plane. For arbitrary collimator angles $\theta_{\text{coll}}$, the \emph{effective} beam size in the collimation plane, $\sigma_{\text{coll}}$ is computed from the
horizontal and vertical sizes as
\begin{equation}
  \sigma_{\text{coll}}=\sqrt{\sigma_x^2\cos(\theta_{\text{coll}})^2
    +\sigma_y^2\sin(\theta_{\text{coll}})^2}~,
\end{equation}
where $\sigma_z$, $z\equiv(x,y)$, is calculated as in \Eref{sz}. The
collimator half gap is calculated as $h=n_\sigma\times\sigma_{\text{coll}}$ and the
jaw positions around the beam position, $x_{\text{beam}}$, are given by
\begin{equation}
  {\text{jaw}}=x_{\text{beam}}\pm n_\sigma\times\sigma_{\text{coll}}~.
\end{equation}
Collimator stepping motors can be driven through arbitrary functions of time. The motion of
collimators around the ring can be synchronized through timing events at the microsecond level \cite{contrIcapeps, collControlsPac09}, which is used to synchronize their motion to the general machine time used for different operational phases. To this end, continuous setting functions
must be generated from the beam-based parameters through scaling rules versus
beam energy and optics.

\begin{figure}
  \centering
  \includegraphics[width=100mm]{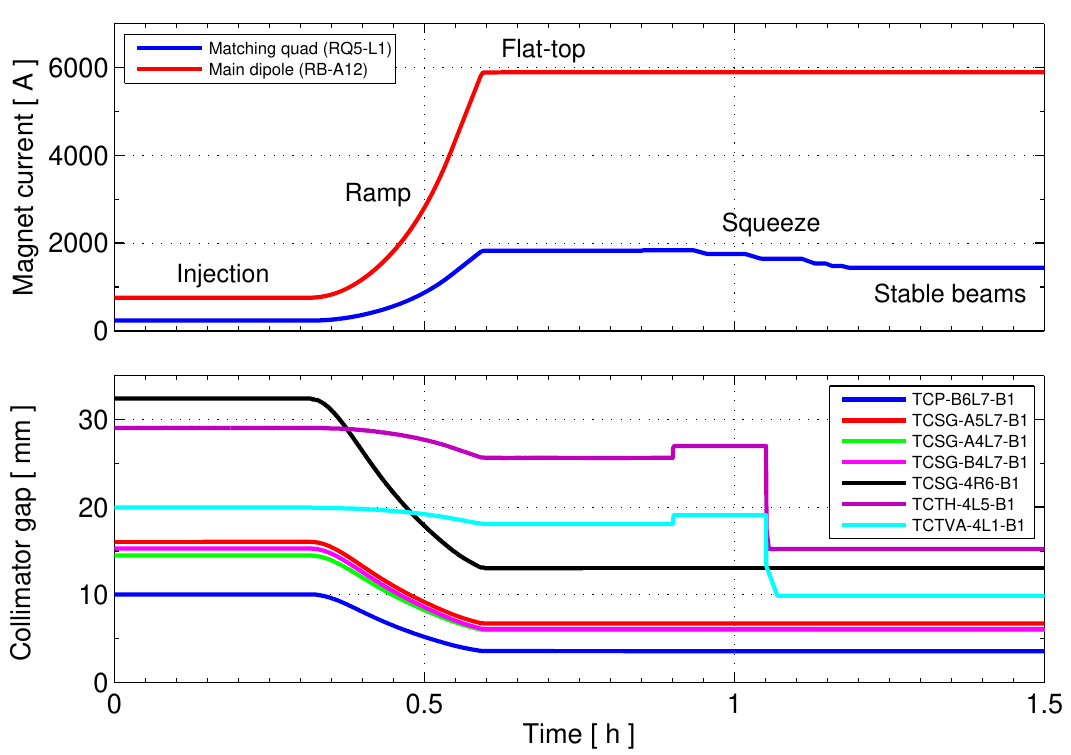}
  \vspace{-0.2cm}
  \caption{Operational cycle for selected collimators for a typical LHC fill.
  Top: Measured magnet currents versus time. Bottom: Collimator gaps versus time.}
  \label{figExFill}
\end{figure}

Let us calculate, for example, the ramp functions, starting from settings
values at injection (`0') and flat-top (`1'). The half gap
during the energy ramp is expressed as a function of the energy:
\begin{equation}
  h(\gamma)=n_\sigma(\gamma)\times\sigma_{\text{coll}}(\gamma)~,
\end{equation}
where $\gamma=\gamma(t)$ is the relativistic gamma function. For the LHC, it
is sufficient to use linear functions in $\gamma$ for $n_\sigma$ and
$\sigma_{\text{coll}}$. A linear interpolation between the beam-based parameters
at injection and flat-top yields:
\begin{equation}
h(\gamma)=
\left[n_{\sigma,0}+
\frac{n_{\sigma,1}-n_{\sigma,0}}{\gamma_1-\gamma_0}(\gamma-\gamma_0)
\right]
\times\frac{1}{\sqrt{\gamma}}
\left[
\frac{
\sqrt{\epsilon_1\beta_1}-\sqrt{\epsilon_0\beta_0}}{\gamma_1-\gamma_0}
(\gamma-\gamma_0)
\right]~.
\end{equation}
%
The beam centre is also expressed as a linear function of $\gamma$ to give
the jaw position as
\begin{equation}
  {\text{jaw}}(\gamma)=\left[x_{\text{beam},0}+\frac{x_{\text{beam},1}-x_{\text{beam}, 0}}
    {\gamma_1-\gamma_0}
    (\gamma-\gamma_0)\right]\pm h(\gamma)~.
\end{equation}
Note that the beam size $\sigma_{\text{coll}}=\sigma_{\text{coll}}(\gamma)$ is also a
function of the optics and therefore might change, typically for the
tertiary collimators in the experimental regions, during the betatron squeeze \cite{srRampSqu}. This notation can be generalized in a straightforward
way by considering functions of $\beta^*$ instead of $\gamma$ for the~parameters involved.
An example of collimator gaps versus time during a full LHC cycle is given in
\Fref{figExFill} (bottom graph), together with the LHC dipole
and matching quadrupole currents, to indicate the~times of the ramp and squeeze phases (top graph). This example refer to a cycle with distinct ramp and squeeze. Settings are generated for the case of combined ramp and squeeze by applying at the same time the $\gamma$ and $\beta$ dependence 

Similar expressions can be derived for the functions of the beam envelope slope. A tilt can be imparted to the collimator jaws -- which are moved by two independent motors -- for example to maintain them parallel to the beam envelope at a certain $n_\sigma$ amplitude. Accounting for the beam divergence from the betatron and off-momentum contributions, for the LHC at top energy, this gives in general small corrections at the level of a few to tens of $\mu$rad that can be neglected. Jaw tilt adjustments are in some case required to compensate alignment errors of the collimators.

The operation of the collimation system is automated by sequences
that are run at every fill, en\-abling operation crews to run
smoothly through the different sets of the cycle settings. The operation mode can
only work thanks to the excellent stability of the LHC orbit and
optics and of the collimator hardware itself. So far, only one beam-based
alignment per year has been required \cite{collPerf}.




\section{Collimator design for high-power accelerators}
\label{design}
Let us review the key features of the collimators for the LHC collimation system. The main design parameters are summarized in Table~\ref{tabParam}. The list emphasizes the challenges in terms of material compatibility to the beams with high damage potential, of heating of components and radiation doses, of impedance. Several aspects need to be accounted for simultaneously to achieve an optimized design. Note that the~design must ensure adequate mechanical stability during jaw position changes and in the~presence of important heat loads. Other aspects related to materials choice to ensure robustness and limited impedance are addressed for example in~\cite{ab}. Details of the final collimator design deployed for the~LHC can be found in Refs. \cite{collDesign, finalColl}. Here, only the main design features are given.

\begin{table}
  \caption{Minimal horizontal and vertical apertures at injection
  ($450\UGeV$) for warm and cold elements}
  \begin{center}
    \begin{tabular}{ll}
      \hline\hline
      \textbf{Parameter} & \textbf{Value} \\ 
      \hline
      High stored beam energy &  $360\UMJ/\text{beam}$\\
      Large transverse energy density& $1\UGJ/\UmmZ^2$\\
      Activation of collimation inserts & 1--15$\UmSv/\UhZ$ \\
      Small spot sizes at high energy & ${\approx}200\Uum$\\
      Collimation close to beam & 6--7$\sigma$\\
      Small collimator gaps & $2.1\Umm$ (at $7\UTeV$)\\
      Big and distributed system & 110 devices, $\approx$500 degrees of freedom\\
      \hline\hline
    \end{tabular}
  \end{center}
  \label{tabParam}
\end{table}


The LHC collimators are high-precision devices that ensure the
correct transverse hierarchy along the $27\Ukm$ long ring with beam sizes as small as $200\Uum$. Each collimator has two jaws, of different lengths and materials,
depending on functionality (Table~\ref{tabList}). Each jaw can be
independently moved by two stepping motors. Key features of the design
are: (1) a
jaw flatness of about $40\Uum$ along the $1\Um$ long active jaw surface;
(2) a surface roughness less than $2\Uum$; (3) a $5\Uum$ positioning resolution;
(4) an~overall setting reproducibility below $20\Uum$ \cite{contrIcapeps};
(5) a minimal gap of $0.5\Umm$; (6) evacuated heat loads of up to $7\UkW$ in
a steady-state regime and of up to $30\UkW$ in transient conditions.

Primary and secondary collimators are made of a robust CFC that is designed to withstand beam impacts without significant
permanent damage for the worst assumed failure cases: impact of a full injection batch of 288 nominal protons and
impact of 8 proton bunches at $7\UTeV$ \cite{collDesign}. Other
collimators made of heavy tungsten alloy or copper, obviously, do not have the
same robustness and are only utilized at larger distances from the circulating beams, where maximum absorption is needed.

\begin{figure}
  \centering
  \includegraphics[width=135mm]{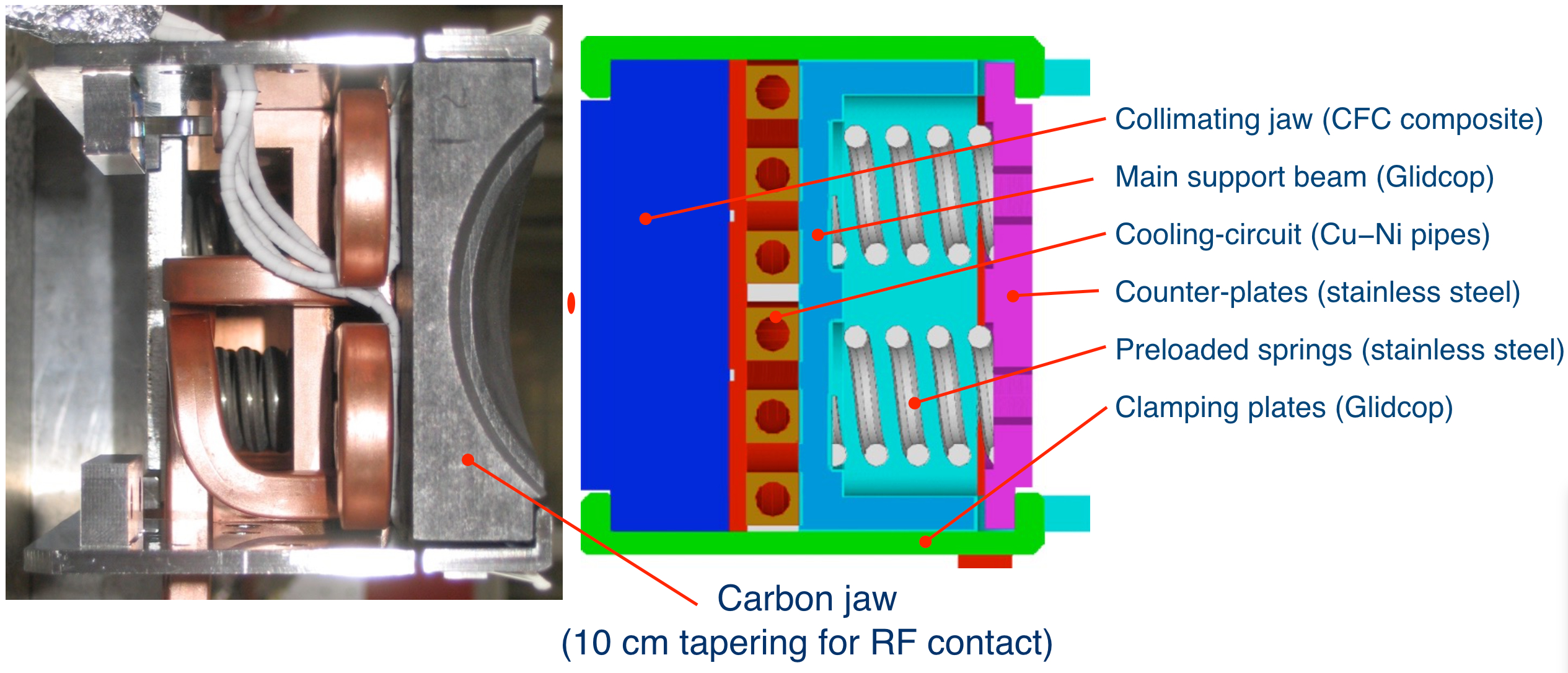}
  \vspace{-0.2cm}
  \caption{Cross-section of the LHC collimator jaws. Left: real prototype. Right: design drawing. The position of the beam is shown by the red
    ellipse, as if the two jaws were those of a horizontal collimator. A
    sandwich structure, with cooling circuits clamped on the
    CFC plate of the active part, is optimized to minimize deformation of the
    structure during steady loss conditions \cite{collDesign}.}
  \label{figXsec}
\end{figure}

\begin{figure}
  \centering
  \includegraphics[width=150mm]{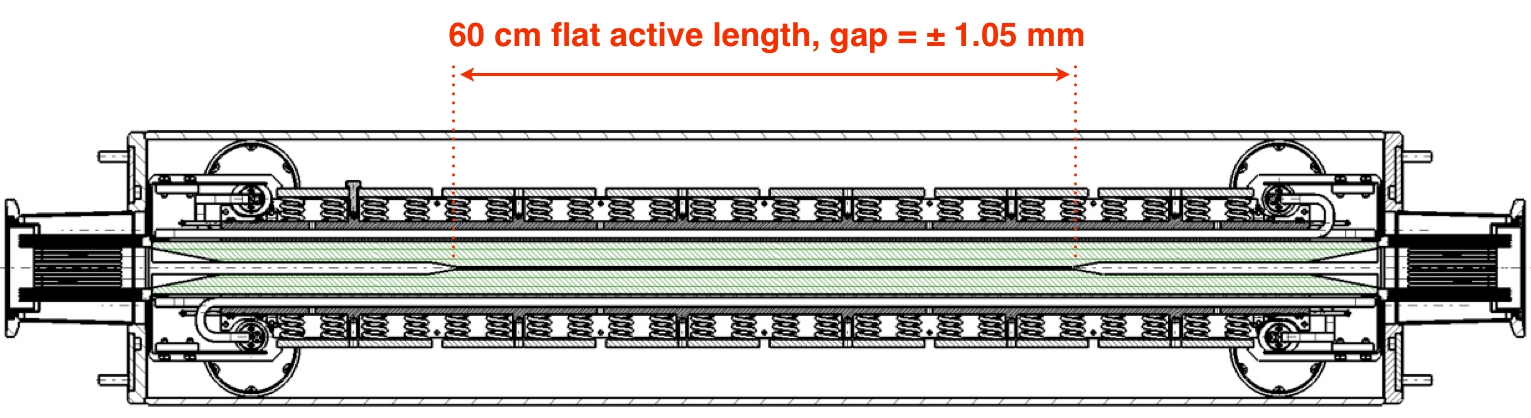}
  \vspace{-0.2cm}
  \caption{Design of the LHC primary collimator. The two jaws can move
  independently, thanks to four stepping motors enabling position and angular
  adjustment with respect to the beam. This design is essentially identical
  to that of the secondary collimator except that the jaws are tapered to an
  effective length of $60\Ucm$ instead of $100\Ucm$.}
  \label{figTCP}
\end{figure}

The cross-section of the primary and secondary collimator jaws, with a $2.5\Ucm$
thick active CFC part and a cooling system underneath, is shown in
\Fref{figXsec}. The design drawing on the right side of the~picture is
compared with a real jaw prototype on the left, built  during the initial
production phases to verify the manufacturing quality.
Two parallel jaws are mounted in the vacuum tank, as shown in \Fref{figTCP}
for a primary collimator. 
In \Fref{figTCP}, the jaws are actually shown in their
operational positions for the~vertical collimator with the tightest gaps, as in \Fref{fig_smallgaps}.

\begin{figure}
  \centering
  \includegraphics[width=60mm]{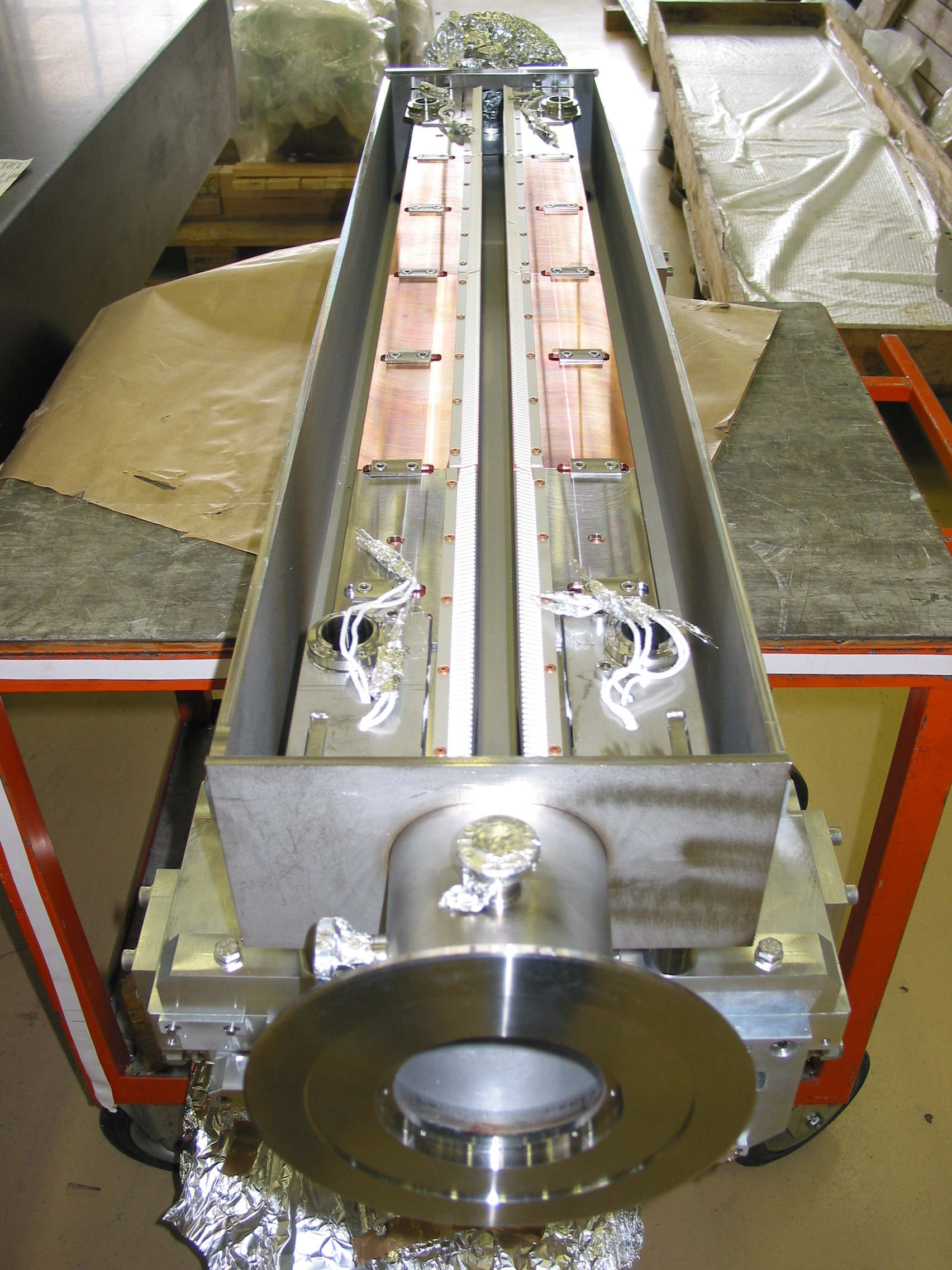}
  \includegraphics[width=80mm]{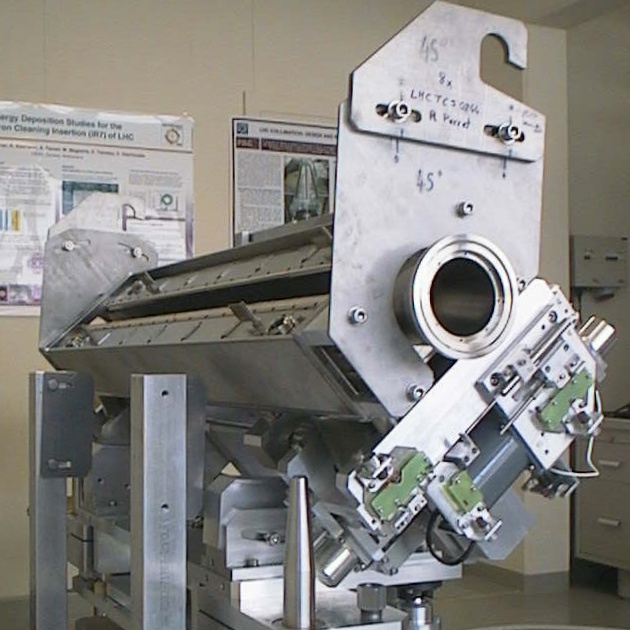}
  \vspace{-0.2cm}
  \caption{Horizontal (left) and skew (right) LHC collimators
    with open tank, showing movable jaws. The support allows assembly
    in the same collimator
    tank of all the required orientations.}
  \label{figCollLab}
\end{figure}

\begin{figure}
  \centering
  \includegraphics[width=80mm]{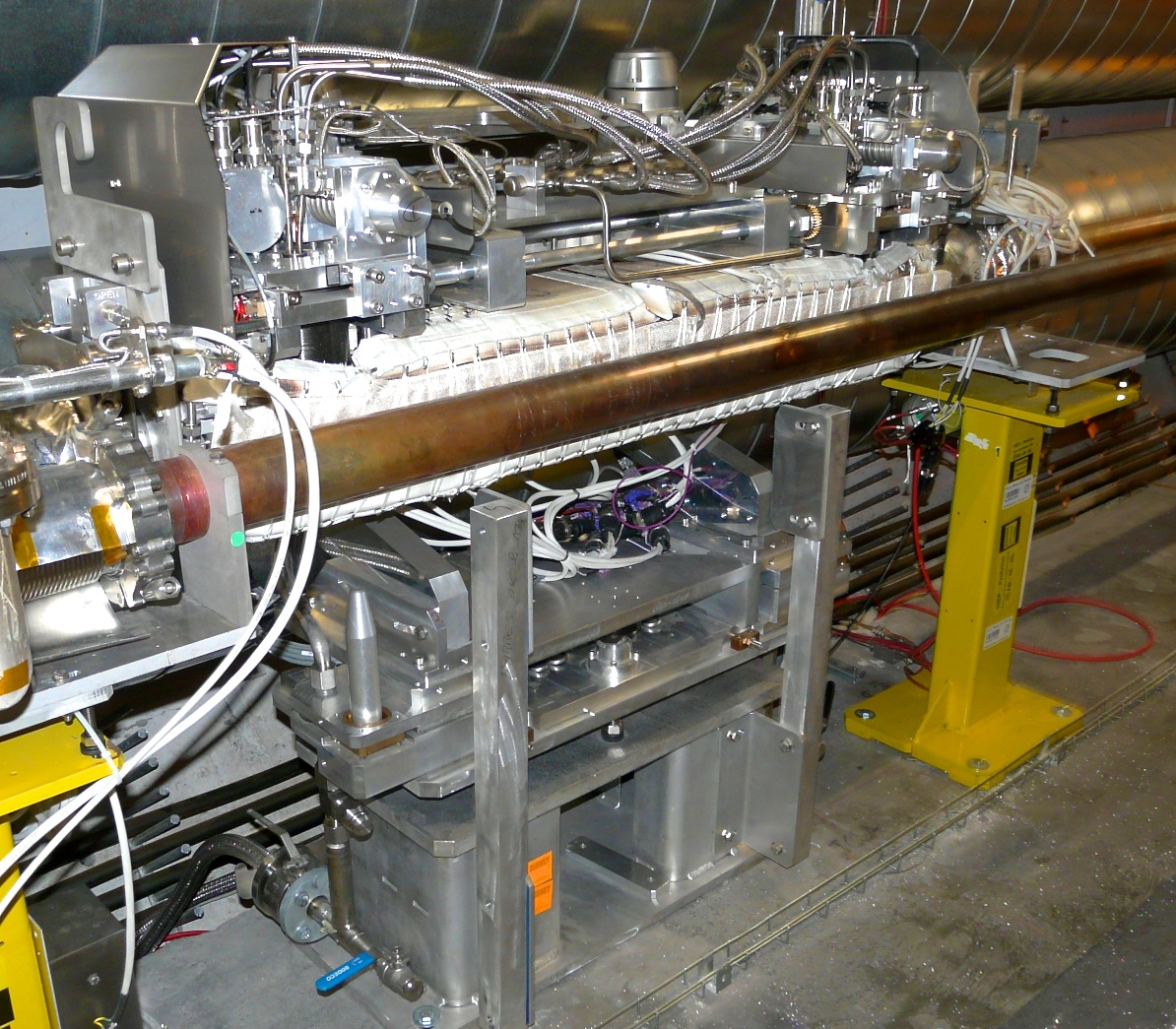}
  \vspace{-0.2cm}
  \caption{Active absorber TCLA.B6R7.B1 as installed in the
    betatron cleaning insert. The stepping motors that control jaw position
    and angle are visible on top of the vacuum tank. The pipe of
    the opposing beam is also shown. }
  \label{figTCLAtunnel}
\end{figure}

\Figure[b]~\ref{figCollLab} shows a horizontal and a $45^{\circ}$ tilted LHC collimator. Their vacuum tank is still open to show the CFC jaws inside. An example of the tunnel installation layout for a IR7 collimator is given in \Fref{figTCLAtunnel}. This is a horizontal TCLA collimator. Notice, next to the collimator, a
yellow support that supports a vacuum pump that is installed next to each collimator. A beam loss monitor, not visible in the photograph, is also
connected to this support, to record losses generated locally when the beam is intercepted by
the collimator jaws.

\begin{figure}
  \centering
  \includegraphics[width=140mm]{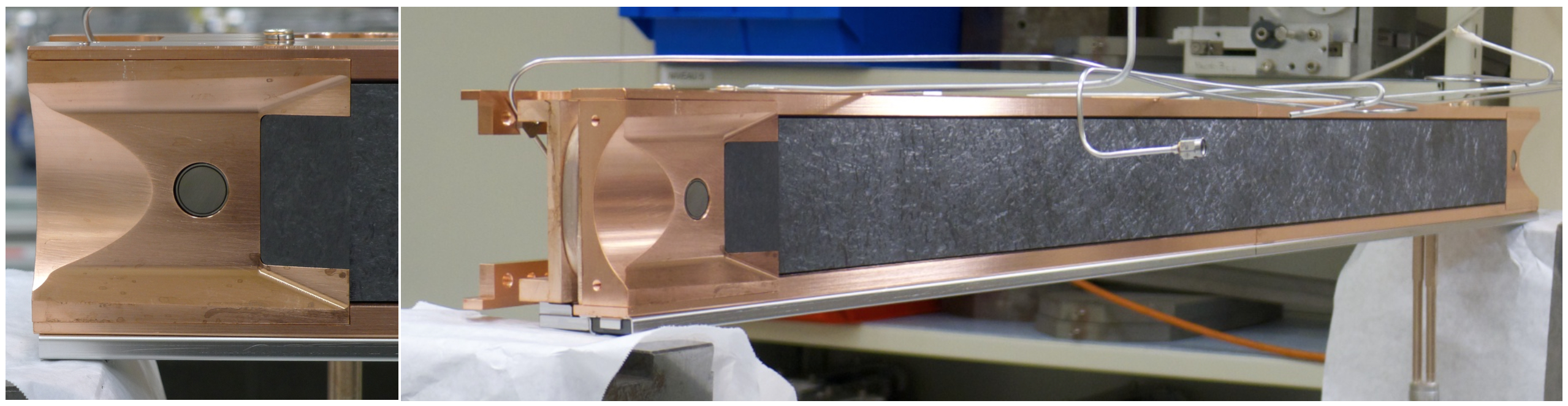}
  \vspace{-0.2cm}
  \caption{New CFC jaw with integrated beam position monitors at each
    extremity for installation in IR6 (see \Fref{figLayout}). A variant
    of this
    design, made with a Glidcop support and tungsten heavy alloy inserts on
    the active jaw part, is used for the new TCTP tertiary collimators in all
    IRs.}
  \label{figTCSP}
\end{figure}

The collimator design has been further improved by adding two beam position monitors on either extremity of each jaw \cite{carra}.
An example of a CFC jaw prototype with this new design is shown in
\Fref{figTCSP}. This feature allows faster collimator alignment as well
as constant monitoring of the beam
orbit at the~collimator, as opposed to the beam-loss-monitor-based alignment that can currently only be performed during dedicated low-intensity commissioning fills. The
beam position monitor buttons will improve collimation performance significantly in terms of
operational efficiency and flexibility, by reducing the~machine time spent on
aligning collimators and the $\beta^*$ reach \cite{collDesign}. The beam-position-monitor-embedded
design is
considered the baseline for future upgraded collimator design.

\section{Cleaning performance of the LHC beam collimation}
\label{lhc-perf}
The cleaning performance of the LHC collimation system is measured by
intentionally generating trans\-verse and off-momentum beam losses while
measuring losses around the ring. This is done with low intensities
circulating in the machine. Individual bunches are excited by adding transverse noise with the
transverse damper. This method allows independent excitation in horizontal and vertical planes, in a~bunch-by-bunch mode that enables performing efficiently several loss maps in single fills. Large losses of
the momentum cleaning can instead be generated by changing the radio frequency. These so-called \emph{loss maps} are
used to validate, empirically, the response of the collimation system in the presence
of high loss rates. This is an essential part of the validation of the LHC
machine protection functionality, as discussed in Ref.~\cite{jw}. In particular,
loss maps are used to verify: (1) that the hierarchy is respected, by checking
that the relative loss rates at the different collimators are in agreement with
predictions or within tolerable levels; (2) that the leakage of losses to
the other machine equipment, in particular superconducting magnets, are
as expected; (3) that the system performance remain stable during long
periods when beam-based alignment is not repeated.

\begin{figure}
  \centering
  \includegraphics[width=143mm]{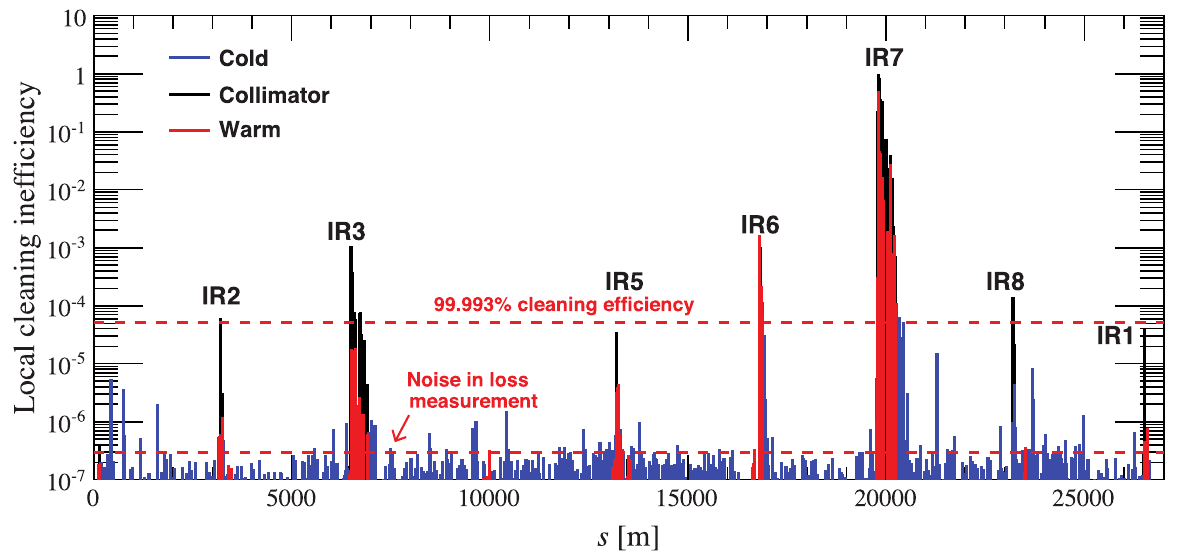}
  \includegraphics[width=143mm]{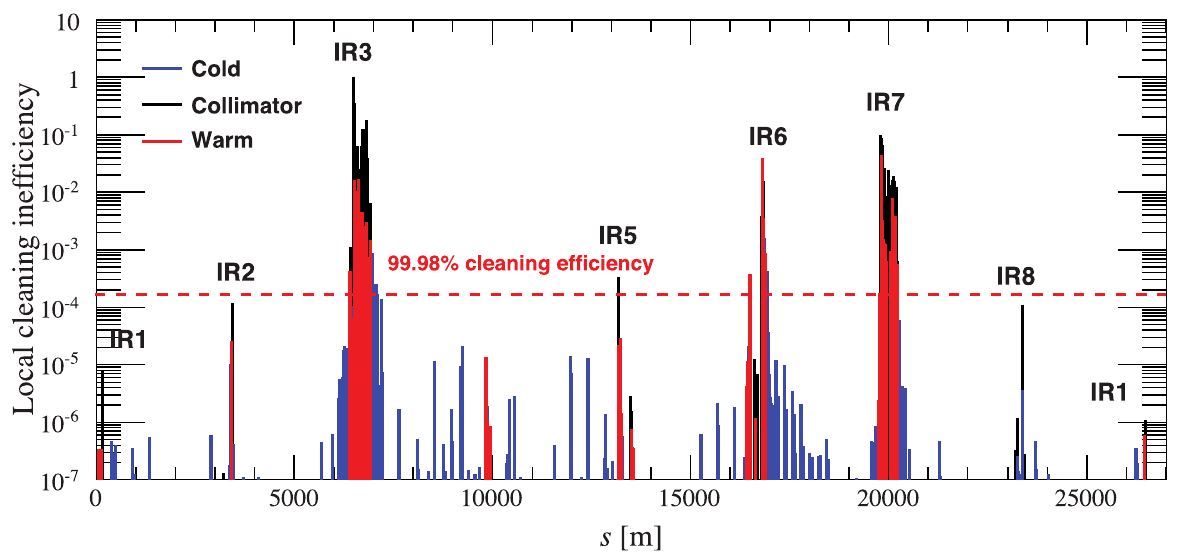}
  \vspace{-0.2cm}
  \caption{Betatron (top) and off-momentum (bottom) loss maps obtained at the LHC at $4\UTeV$ with beams squeezed to $60\Ucm$ in IR1 (ATLAS) and IR5 (CMS), showing the beam losses recorded at about 4000 beam loss monitors around the ring,
    normalized to the highest measured signal.
    Betatron losses are generated in IR7 by adding noise to the kickers of     the transverse damper of clockwise beam 1. IR3 losses are generated by     changing the radio frequency until the full beam is intercepted by the IR3
    TCP. Both beams are excited at the same time as their frequencies are     synchronized. {\it Courtesy of B.~Salvachua}~\cite{collPerf}.}
  \label{figLM}
\end{figure}

\begin{figure}
  \centering
  \includegraphics[width=120mm]{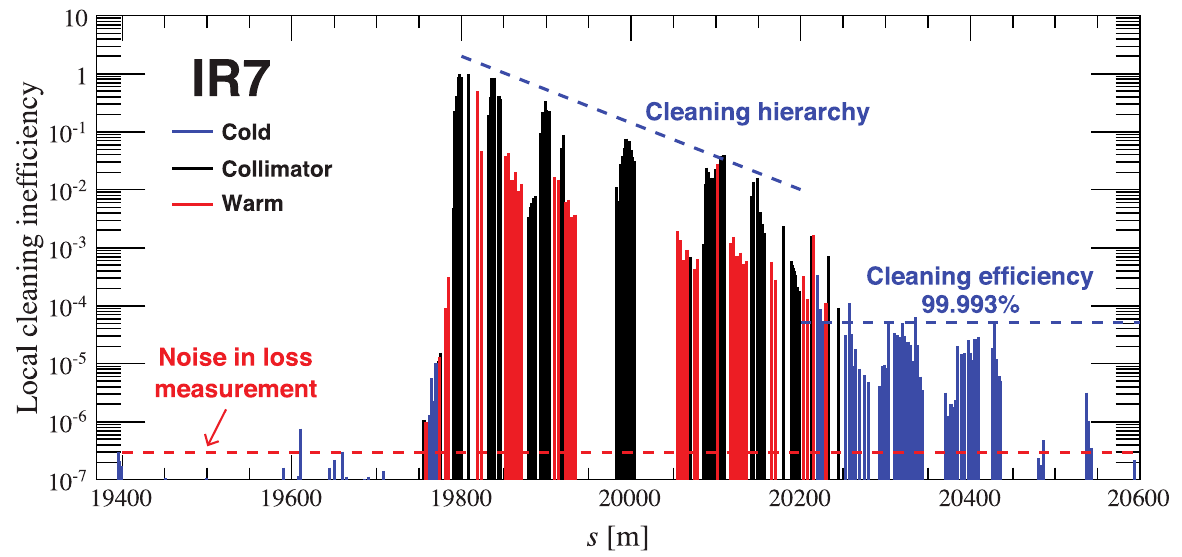}
  \vspace{-0.2cm}
  \caption{Enlargement of the top graph of \Fref{figLM}, showing details of
    losses in IR7. The limiting location for betatron cleaning is given
    by the losses on the cold magnets in the dispersion suppressor immediately
    downstream of IR7.}
\label{figLMz}
\end{figure}


Examples of betatron and off-momentum loss maps are shown in \Fref{figLM}.\footnote{The present LHC cycle features significantly greater complexity compared to the simpler cases discussed in this lecture for illustration. Various techniques for $\beta^*$ and beam-orbit control (separation and crossing) are used during collisions to optimize luminosity. The validation of LHC loss maps has correspondingly grown in complexity, typically requiring several tens of loss maps. However, the fundamental approach remains the same as that used for years with simpler operational cycles.}
These maps were recorded in 2012 at $4\UTeV$, with beams squeezed to $60\Ucm$ in IR1 and IR5. At the LHC, beam losses are recorded by about 3600 beam loss monitors (BLMs) distributed around the ring~\cite{bd}. To estimate the cleaning inefficiency, losses at each monitor are normalized to the highest measured signal, \ie those next to the primary collimators. This is shown in \Fref{figLM} as a function of the longitudinal coordinate $s$. Inefficiencies below ${\sim}10^{-4}$ were achieved. In all IRs, the largest losses are recorded at the collimators (black bars), as expected. The cold locations with the highest losses are the dispersion suppressors downstream of the cleaning inserts, as predicted in simulations (see \Fref{fig_cl}).

A key ingredient for operational performance is that critical LHC machine parameters — such as orbit, optics, and the mechanical stability of collimator settings — are highly reproducible. It has so far been demonstrated that the outstanding performance shown above can be maintained with just one beam-based alignment per year of the ring collimators. In the present operational mode, the alignment and establishment of reference settings are carried out annually during initial beam commissioning. Collimation cleaning is then regularly verified with periodic loss maps to detect any early degradation in performance.

The IR7 losses are shown in \Fref{figLMz}. The limiting locations with the poorest cleaning correspond to the dispersion suppressor magnets on either side of IR7 (blue peaks on the right side of IR7 for beam~1 in the example shown). A cleaning efficiency above 99.993~\% was achieved, with the worst performance found at a few isolated peaks. Elsewhere, the rest of the cold machine experiences losses more than one order of magnitude smaller — \ie close to the noise level of the beam loss monitor (BLM) system. 

In simulations, losses are sampled using $10\Ucm$ bins by counting the number of beam particles hitting the aperture. In measurements, losses are recorded at discrete BLM locations (about 3600 monitors distributed around the ring). The BLMs measure the flux of ionizing particles within their volume. Clearly, these two quantities cannot be directly compared without additional energy-deposition simulations starting from the multiturn loss patterns. A detailed discussion of this aspect is beyond the scope of this lecture. It suffices to say that the agreement between simulations and measurements is good~\cite{Bruce}. Current observations~\cite{Redaelli:2020mld} indicate that the LHC collimation system fulfills the design requirements and is compatible with the challenges of the HL-LHC, at least for operation with proton beams.



\section{Concluding remarks}
\label{concl}

This lecture provides an overview of the key design requirements for high-performance multistage collimation systems in present and future particle accelerators. The LHC collimation system is presented in detail as a case study, offering a comprehensive view of the state of the art in beam-halo cleaning performance. The concepts discussed can be applied directly to other accelerators with less demanding requirements and form a basis for designing future machines, such as the Future Circular Collider (FCC). The design criteria outlined here are actively used for both the hadron-hadron and electron-positron FCC designs.

Several important aspects of beam collimation are beyond the scope of this lecture. Advanced topics, actively pursued in current research, can be explored in the references provided. These include HL-LHC upgrades~\cite{Redaelli:2020mld}, such as local cleaning in cold magnets, crystal collimation of ion beams~\cite{Redaelli:2025tpn}, and hollow electron lenses for active halo control~\cite{Redaelli:2021brx}, as well as the nonlinear collimation scheme deployed at SuperKEKB~\cite{Terui:2024rdo}. Interested readers are encouraged to consult these references for further details.

\section*{Acknowledgments}
The material presented here is the result of the work carried out during several years by many different people. It was a pleasure for me to collect it and presented on behalf of the LHC collimation team. Past and present
team members are gratefully acknowledged, in particular the ones who provided material included in this manuscript. The CERN Accelerator School (CAS) team is also kindly acknowledged for inviting me to prepare this lecture and for their support in the preparation of these proceedings.

\end{document}